\RequirePackage{fix-cm}
\documentclass[smallextended]{svjour3}       
\usepackage{chicago}
\usepackage{algorithm}
\usepackage{algorithmic}

\usepackage{mathptmx}      
\usepackage[english]{babel}
\usepackage[pdftex]{graphicx}
\usepackage{hyperref}
\usepackage{color}
\usepackage[utf8]{inputenc}
\hypersetup{
pdfauthor=Havemann et al.
pdftitle=Memetic search for overlapping topics,
colorlinks=true,
linkcolor=black,
anchorcolor=black,
citecolor=blue,
urlcolor=black,
menucolor=black
}

\journalname{Scientometrics}
\begin{document}

\title{Memetic search for overlapping topics \\based on a local evaluation of link communities}

\titlerunning{Memetic Search for Overlapping Topics}        

\author{Frank Havemann \and Jochen Gl\"{a}ser \and Michael~Heinz}


\institute{Frank Havemann and Michael Heinz \at
              Institut f\"{u}r Bibliotheks- und Informationswissenschaft, Humboldt-Universit\"at zu Berlin, D~10099 Berlin, Dorotheenstr. 26, Germany \\
              Tel.: +49-30-20934228\\
              \email{frank.havemann@ibi.hu-berlin.de}           
           \and
           Jochen Gl\"{a}ser \at
Center for Technology and Society, 
TU Berlin (Germany)
}

\date{Received: date / Accepted: date}

\maketitle

\begin{abstract}
\sloppypar
In spite of recent advances in field delineation methods, bibliometri\-cians still don't know the extent to which their topic detection algorithms reconstruct `ground truths', i.e.\ thematic structures in the scientific literature. In this paper, we demonstrate a new approach to the delineation of thematic structures that attempts to match the algorithm to theoretically derived and empirically observed properties all thematic structures have in common.
We cluster citation links rather than publication nodes, use predominantly local information and search for communities of links starting from seed subgraphs in order to allow for pervasive overlaps of topics. 
We evaluate sets of links with a new cost function and assume that local minima in the cost landscape correspond to link communities. 
Because this cost landscape has many local minima we define a valid community as the community with the lowest minimum within a certain range. Since finding all valid communities is impossible for large networks, we designed 
a memetic algorithm that combines probabilistic evolutionary strategies with deterministic local searches. 
We apply our approach to a network of about 15,000 Astronomy \& Astrophysics papers published 2010 and their cited sources, and to a network of about 100,000 Astronomy \& Astrophysics papers (published 2003--2010) which are linked through direct citations.
\end{abstract}

\section{Introduction}
\label{intro}
The identification of thematic structures (topics or fields) in sets of papers is one of the recurrent problems of bibliometrics. It was deemed one of the challenges of bibliometrics by \citeN{vanRaan1996advanced} and is still considered as such despite the significant progress and a plethora of methods available. Major developments since van Raan's paper include approaches that cluster the whole Web of Science based on journal-to-journal citations \cite{leydesdorff2004clusters,leydesdorff2009global,leydesdorff_interactive_2012}, co-citations, or direct citations \cite{boyack2005mapping,boyack2010co-citation,klavans2011using,waltman_new_2012}, the advance of hybrid approaches that combine citation-based and term-based techniques~\cite{Glenisson2005combining,Glaenzel2015ISSI}, and term-based probabilistic methods (topic modelling, cf.\ \citeNP{yau_clustering_2014}). 
 However, in spite of all the methodological progress we still don't know whether the groups of papers found by these methods represent the knowledge structures they are intended to identify. It is still not clear which, if any, `ground truth' about the set of papers is reflected by the groups of papers.

 This uncertainty also applies to community detection, which represents one group of methods that can be used for the reconstruction of thematic structures in sets of papers. The recent discussion about the ability of community detection algorithms to reconstruct ground truths that are reflected in metadata \shortcite{hric_community_2014,peel_ground_2016} illustrates the point. Metadata are likely to reflect the various ground truths differently. Peel et al.\ identified several reasons why the division of a network might not correlate with metadata (p.\,2), and stated further that a network ``can result from multiple, distinct generative processes, each with its own ground truth'' (p.\,3). Applied to the reconstruction of thematic structures from networks of papers, this means that although we know the metadata about papers and the links in a network to be related to the ground truths we intend to reconstruct, it is still difficult to tell which of these ground truths (thematic structures), if any, a community detection algorithm reconstructs.

 In this paper, we propose a new approach to this problem. We attempt to increase the likelihood that relevant thematic structures are reconstructed from the network of papers by using an algorithm that fits structural properties all the ground truths we are interested in have in common. 
We start from a theoretical discussion  
of thematic structures in science and derive principles of clustering from theoretically deduced and empirically observed properties of topics. 
Applying these principles, we cluster citation links rather than publication nodes, use predominantly local information, and search for communities of links starting from seed subgraphs in order to allow for pervasive overlaps of topics. 

We evaluate sets of links with a new cost function and assume that local minima in the cost landscape correspond to link communities. 
\label{f:why memetics} 
Since identifying all these communities in a large network is impossible in reasonable periods, we apply 
a memetic algorithm that combines nondeterministic evolutionary strategies with deterministic local searches in a cost (or fitness) landscape \cite{neri2012handbook}. We demonstrate and test our approach by applying it to two networks. The first network consists of about 15,000 Astronomy \& Astrophysics papers (published 2010) and their cited sources (see \citeNP{Haveman2015ISSI}  for preliminary results). The second network consists of about 100,000 Astronomy \& Astrophysics papers (published 2003--2010) which are linked through direct citations. 

\section{Theoretical considerations}

We start from the sociological insight that topics are actively constructed by researchers in their joint production of scientific knowledge. We define a topic as \textit{a focus on theoretical, methodological or empirical knowledge that is shared by a number of researchers and thereby provides these researchers with a joint frame of reference for the formulation of problems, the selection of methods or objects, the organisation of empirical data, or the interpretation of data}
(on the social ordering of research by knowledge see 
\citeNP{glaser_wissenschaftliche_2006}). This definition resonates with \citeANP{whitley1974cognitive}'s \citeyear{whitley1974cognitive} description of research areas but abandons the assumption that topics form a hierarchy. 
The only demand the definition makes is that some scientific knowledge is perceived similarly by researchers and influences their decisions.

From the definition follows that there is no structural or functional difference between a topic and a field,
as was suggested by sociologists who considered the hierarchical order of thematic and social structures, e.g.\ \citeN{whitley1974cognitive} and  \citeN{Chubin1976conceptualization}. 
We consider both topics and fields as shared perspectives on knowledge.

All topics and fields emerge from coinciding autonomous interpretations and uses of knowledge by researchers (see e.g.\ the case studies of emerging topics/fields discussed by \citeNP{edge1976astronomy}, pp. 350--402). While individual researchers may launch topics and advocate them, the latter's content and fate depend on the ways in which they are used by others.

This constructivist 
approach explains three properties of topics that have consequences for bibliometric methods:
\begin{enumerate}
\item
Topics are local in the sense that they are primarily topics to the researchers whose decisions are influenced by and who contribute to them. 
Methods for topic identification using only local information reconstruct this insider perspective, while methods using also global information construct a compromise between insider and outsider perspectives on topics.

\item 
Topics can have any `size' (however measured) between the smallest topics (that just concern very few researchers) and very large thematic structures (fields or even themes cutting across several fields). This fractal nature of knowledge has been described by \citeN{vanRaan1991fgi} and \citeN{katz1999sss}. Methods aimed at recovering this structure should not be biased against any particular topic size.
\item
Given the multiple objects of knowledge that can serve as common reference for researchers, topics inevitably overlap. Publications commonly contain several knowledge claims, which are likely to address different topics \cite{cozzens_comparing_1985,amsterdamska_citations:_1989}. 
Adjusting to this property of topics means taking into account that 
bibliometric objects like terms, publications, authors, journals, and citation links are likely to belong to several topics simultaneously. 
Consequently, the reconstruction of overlapping topics in publication networks requires that subgraphs may overlap pervasively, i.e.\ not only in their boundaries. 

\end{enumerate}
The local construction of topics, their varying size and pervasive overlaps make it likely that
topics form a poly-hierarchy 
i.e.\ a hierarchy where a smaller topic can be a subtopic of two or more larger topics that have no hierarchical subtopic relation (see \citeN[239--241]{Healey1986esm}  for the observation of a poly-hierarchy of topics based on keywords).

An attempt to reconstruct topics with the above-described properties needs to apply four principles.  
\textit{First}, clustering
should be based on the least heterogeneous property of a paper available to bibliometrics, namely the link between a publication and its cited source. 
A citation often links a paper to one specific knowledge claim made in the cited source. Although a publication may be cited in a paper without reference to a specific knowledge claim, or may be cited several times for the use of different knowledge claims, links to cited sources are likely to reflect only one topic in most cases.\footnote{The same argument can be made for a term used in one paper. Exploiting this property alone and in combination with links to cited sources is a task for future work.}
\textit{Second}, clustering should utilise mainly local information when the `producer perspective' on topics is to be reconstructed. \textit{Third}, an algorithm should enable pervasive overlaps of subgraphs in order to reconstruct overlapping
thematic structures.
 \textit{Fourth}, an algorithm should not be biased against any particular topic size.

The construction of overlapping communities is by now a well-known and frequently addressed problem of network analysis \cite{fortunato2010community,Xie:2013:OCD,amelio_overlapping_2014}.
Constructing overlapping communities of nodes by clustering links has been proposed by \citeN{evans2009line} and by \citeN{ahn2009link}. 
The use of local rather than global information to reconstruct community structures of networks is also well established \cite{clauset2005flc,lancichinetti2009detecting,havemann2011identification,zhang_detecting_2015}. 
However, as reviews of algorithms indicate \shortcite{fortunato2010community,Xie:2013:OCD,amelio_overlapping_2014}, link clustering, 
the use of local information and the construction of 
pervasively overlapping communities have not yet been applied together, possibly because the task for which this is necessary has not yet arisen.

\section{Method}
We operationalise `topic' as a community 
in a citation network of papers, i.e.\ as a 
cohesive subgraph that is well separated from the rest of the network 
 \cite[p.\ 83]{fortunato2010community}. 
\label{def:comm} 
Hierarchical and poly-hierarchical organisation inevitably reduce cohesion because larger communities include boundaries between well separated sub-communities.

From the theoretical considerations follows that links rather than nodes should be clustered for the reconstruction of topics. A link community can be defined as a set of links that is well separated from the rest of the graph and also relatively well connected internally.
The boundary of a community of
links consists of nodes that have links which belong to the community and links that do not belong. This makes it possible
to determine  the degree to which a paper addresses a topic by calculating the proportion of its citation links belonging to the 
respective link community.

Although a citation link is directed---the citing and cited paper cannot be inter\-changed---we can treat citation links as undirected because for the clustering only the content of the link matters. This content is jointly defined by the citing and the cited paper.

We now describe the local cost function for the evaluation of subgraphs (and the identification of link communities), the memetic algorithm in which the cost function is applied, and the 
experimental setup in which the algorithm was applied.

\subsection{Cost function for evaluation of link sets}
\label{sec:Psi}

Link clustering has been introduced by \citeN{evans2009line} and \citeN{ahn2009link}.
Following a suggestion by \citeN{evans2009line} we introduce a 
local cost function $\Psi(L)$  of link set $L$ 
which is based on external and also on internal 
connectivity of $L$ and includes a size normalisation that accounts for the finite size of the network. 

Each link  set $L$ defines  
a subgraph of the network. The subgraph's node set contains all nodes attached to links in $L$. In this sense we use the term \textit{subgraph} also for a link set $L$.

The internal degree $k_i^\mathrm{in}(L)$ of node $i$ is defined as the number of links in $L$ attached to $i$. Its external degree $k_i^\mathrm{out}(L)$ is the number of its other links and is obtained by subtracting its internal from its total degree: $k_i^\mathrm{out}(L) =k_i - k_i^\mathrm{in}(L)$. External degrees $k_i^\mathrm{out}(L)$ are
weighted with the subgraph membership-grade 
$k_i^\mathrm{in}(L)/k_i$ 
of boundary node $i$ to obtain a measure of external connectivity of link set $L$:
\begin{equation}
\sigma(L)    = \sum_{i = 1}^n\frac{k_i^\mathrm{in}(L)  k_i^\mathrm{out}(L)}{k_i},
\label{eq:sigma}
\end{equation} 
where $n$ is the number of all nodes.\footnote{The sum runs through all nodes but only  boundary nodes of $L$ contribute to it because for inner nodes of $L$  we have $k_i^\mathrm{out}(L) = 0$ and for nodes not attached to links in $L$ we have $k_i^\mathrm{in}(L)=0$.} 
Each term in the sum equals the electrical conductance between external and internal nodes connected through node $i$ if we identify the links with electrical resistors of  conductance equal to 1.\footnote{Applying Kirchhoff's laws we obtain total resistance of all links of node $i$  as $1/k_i^\mathrm{in}(L) +1/k_i^\mathrm{out}(L) = [k_i^\mathrm{in}(L)+  k_i^\mathrm{out}(L)] / [k_i^\mathrm{in}(L)  k_i^\mathrm{out}(L)]=k_i/[k_i^\mathrm{in}(L)  k_i^\mathrm{out}(L)]$.} Alternatively, external connectivity $\sigma(L)$ can be justified by translating external connectivity in the properly weighted line graph of the network into terms of the original network \cite{havemann_detecting_2015}.

A simple size normalisation that accounts for the finite size of the network is achieved by adapting the \textit{normalised cut}---suggested by \citeN{shi_normalized_2000} for node communities---to link communities, which leads us to the cost function \textit{normalised node-cut} $\Psi(L)$:
\begin{equation}
\Psi(L)= \frac{\sigma(L)}{k_\mathrm{in}(L)(1-k_\mathrm{in}(L)/2m)},
\label{eq:Psi}
\end{equation} 
where $m$ is the number of all links and $k_\mathrm{in}(L)$ is the sum of all internal degrees $k_i^\mathrm{in}(L)$---cf.\ \shortciteN[p.\ 4]{havemann_detecting_2015}. $\Psi(L)$  is a non-negative function with a maximum value of one.

We construct a cost landscape where each place corresponds to a link set $L$. A place's height is given by cost $\Psi(L)$. Its neighbouring places can be reached by adding a link to $L$ or by removing a link from $L$.    
A local minimum in the cost landscape corresponds to a link set $L$ with $\Psi(L)$ smaller than $\Psi$-values of all link sets surrounding $L$ in the cost landscape. 
Thus, it corresponds to a well separated link set---compared to its surrounding link sets. 
This is why we can define link communities as link sets corresponding to local minima of function $\Psi(L)$. 
In accordance with our understanding of communities as cohesive subgraphs (see above, p.\ \pageref{def:comm}), we also require that link communities must be connected link sets  {\shortcite[p.\,5]{havemann_detecting_2015}}. 

\begin{figure}[!t] 
\includegraphics[width=2in]{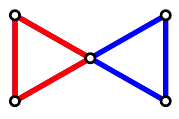} 
\caption{Bow-tie graph} \label{bow-tie} 
\end{figure}

As a simple example, we determined the $\Psi$-landscape of the bow-tie graph (Figure  \ref{bow-tie}).\footnote{cf.\ \citeN{evans2009line} and for calculations see  \shortciteN[Appendix]{havemann_detecting_2015}} We expect a cut through the central node to be the best division in two link communities (the two triangles). Indeed, the landscape has two minima with 
$\Psi = 1 / 3$, which both correspond to a cut through the central node that partitions the graph in two triangles.\footnote{The central node is the boundary node for both triangles with equal internal and external degree of 2. This gives $\sigma(L) = 1$ and $\Psi = 1 / 3$.
All other partitions cut through two nodes and have higher values of $\Psi$.}

The evaluation function can be used to identify link communities by finding local minima in the cost landscape. Since the cost landscape is often very rough---has many local minima that sometimes correspond to very similar communities---it is reasonable to sharpen the criterion for a solution by widening the set of subgraphs we compare with each other. A community has a better $\Psi$-value than all its direct neighbours in the landscape
but for a \textit{valid} community we demand that it has to be better than all link sets in a wider environment. 
The size of this environment is determined by setting a minimum distance between a valid community and any  link set with a lower $\Psi$-value but we must exclude unconnected link sets from the comparison.\footnote{We only compare connected link sets because for a  community $L$ one can construct an unconnected link set $L \cup L_s$ with $\Psi(L \cup L_s) < \Psi(L)$ (which makes $L$ invalid) by adding a small link set $L_s$ to $L$ 
if $\Psi(L_s)$ is small enough and $L_s$ has no connection to $L$.\label{conn}}
We therefore define the \textit{range} $R(L)$ of a community $L$ as the distance in the cost landscape to the nearest 
connected link set 
with a lower $\Psi$-value diminished by one. 
That means, $R(L)$ is the radius of a community's largest environment in the cost landscape in which no connected link set with lower cost exists. 

With the notion of range we  can define a community $L$ as \textit{valid} if it has
a range $R(L) \ge R_\mathrm{min}$. Minimum range $R_\mathrm{min}$ determines the 
\label{resolution}
resolution of the algorithm because any
two valid communities differ by at least $R_\mathrm{min}$ links. 
The algorithm's resolution can be made size dependent by specifying it in relative terms, as we did in our experiments when setting a valid community's minimum range at one third of its size:
$R(L) \ge R_\mathrm{min} = |L|/3.$

Since we are ultimately interested in topics that are addressed by papers, we transform link communities into paper communities by assigning papers fully to a topic if all their citation links belong to the link community and assigning papers fractionally to a topic according to the proportion of their links that belongs to the link community.

\subsection{Memetic algorithm for finding link communities in the cost landscape}
\label{sec:mem}

\begin{algorithm}[!b]
\caption{Pseudocode of memetic evolution for one adapted seed}
\label{pseudcode-mem-evol}
 \begin{algorithmic}
 \STATE     
\STATE  \textbf{initialise} population \textbf{P} by
\STATE \textbf{mutating} the adapted seed with high variance several times and \textbf{adapting} mutants
\WHILE {the best community is not too old}
  \STATE  \textbf{mutate} the best community with low variance and \textbf{adapt} the mutants
	\IF {an adapted mutant is new and its cost is lower than highest cost}
   \STATE  \textbf{add} it to population \textbf{P} 
  \ENDIF 
  \STATE \textbf{cross}  best community with randomly chosen communities and \textbf{adapt} the offspring
  \IF {adapted offspring is new and its cost is lower than highest cost}
 \STATE  \textbf{add} it to population \textbf{P} 
\ENDIF 

 \STATE  \textbf{select} the best communities so that the population size remains constant 

\IF {there is no better best community for some generations and innovation rate is low} 
\STATE  \textbf{renew} the population by mutating the best community with high variance and \textbf{adapt} mutants 
 \STATE  \textbf{select} the best communities so that the population size remains constant 

 \ENDIF 
   \ENDWHILE
 \STATE
 \end{algorithmic}
\end{algorithm}

The cost function $\Psi$ is used in a clustering algorithm that looks for minima in the cost landscape starting from seed subgraphs of different sizes. 

The task of finding communities in large networks is always very complex and requires the use of heuristics. We chose a memetic algorithm that  combines non-deterministic evolution with a deterministic local search in the cost landscape \cite{neri2012handbook}. In our algorithm, a population of communities is randomly initialised and evolves because 
the genetic operators of crossover, mutation, and selection are repeatedly applied. 
Since the non-deterministic operators do not necessarily lead to communities, each crossover and mutation is followed by a local search, which we call adaptation,
cf.\ Algorithm \ref{pseudcode-mem-evol}. 
The genetic operators and local search procedures
are described in Appendix \ref{app.mem}, p.\ \pageref{app.mem}.
The size of the population is kept constant.

Since the evolutions of community populations are independent from each other, the final communities obtained in different evolutions can overlap pervasively. An evolution ends if the $\Psi$-value of the best community does not improve anymore regardless of community size. By preventing a size bias of the algorithm and enabling pervasive overlaps, the independent construction of communities fulfils two more principles derived from our definition of a topic.

Evolutionary algorithms have already been used for identifying communities in networks \cite[p.\ 106]{fortunato2010community}.
Some authors have even applied evolutionary algorithms to link clustering but all used global evaluation functions \cite{pizzuti_overlapped_2009,li_discovering_2013,shi_link_2013}.
Memetic evolutionary algorithms have also been applied to reconstruct communities but only for node clustering and only with global evaluation functions \cite{gong_memetic_2011,pizzuti_boosting_2012,gach_memetic_2012,ma_multi-level_2014}.

\subsection{Experimental setup}
\label{sec: exp}
The overall procedure is described by  Algorithm \ref{pseudcode-exp}. We apply a node-wise local search inside the memetic algorithm and a link-wise local search at the end (s.\ Appendix \ref{app.mem}, p.\ \pageref{app.mem}).

\begin{algorithm}[!b]
\caption{Pseudocode of the whole procedure}
\label{pseudcode-exp}

 \begin{algorithmic}
 \STATE      
\WHILE {coverage with  valid communities is improving}
  \STATE  \textbf{select} a seed subgraph and \textbf{adapt} it by a node-wise local search
  \STATE \textbf{run Algorithm \ref{pseudcode-mem-evol}} up to five times with fixed mutation variance 
  \STATE  \textbf{select} best result as seed community 
  \STATE  \textbf{run Algorithm \ref{pseudcode-mem-evol}} up to ten times with decreasing mutation variance
  \STATE \textbf{adapt} the best result by link-wise local search 
  \IF {the adapted result is an unconnected subgraph}
    \STATE  \textbf{adapt} components by link-wise local search
  \ENDIF
   \STATE \textbf{test validity} of all previous results 
    \STATE \textbf{determine coverage} by valid communities 
\ENDWHILE
 \STATE \textbf{select} communities with size between one quarter and three quarters of all links
\STATE \textbf{adapt} components of their complements by link-wise local search
 \STATE \textbf{test validity} of all previous results
 \STATE 
 \end{algorithmic}
\end{algorithm}

After the experiments were completed, we found that one sub-routine (the node-wise local-search within the memetic evolution) calculated the cost function incorrectly.%
\footnote{ Instead of the function $\Psi$ defined in equation \ref{eq:Psi}, the following function was implemented: $(\sigma(L)/k_\mathrm{in}(L))(1-k_\mathrm{in}(L)/2m)$.} 
Since all other sub-routines using the cost function calculated it correctly, the error affected primarily the efficiency of the algorithm rather than the validity of its results. However, the bug created a tendency for the memetic search to overlook some local minima when analysing large sub-graphs. We therefore repeated the experiments for all seeds that did not lead to communities in the first two experiments 
and for several other large seeds. This re-run identified some further large valid communities. In other cases, the memetic algorithm only led to variants of communities found before, thereby confirming the limited impact of the bug. 

Because the algorithm searches randomly, communities can be improved by any new experiments or new valid communities can be found. In this sense, for larger networks our algorithm never comes to a definite end.

The experimental setup includes setting a resolution parameter $r < 1$ 
which determines a valid community's minimum range  according to 
$R_\mathrm{min}(L) = r|L|.$ We choose $r=1/3$.

We also decided to consider communities of more than three quarters of all links as invalid because their range exceeds the whole network, and their validity cannot be determined. 

\paragraph{Initialising populations from seed subgraphs:}
Since topics vary in size the algorithm should start from differently sized seed subgraphs. Seed subgraphs should also be located in all regions of the network. Owing to the randomness of evolution the choice of seed graphs is unlikely to affect the  results. However, it is likely to effect the efficiency of the algorithm.
In our experiments, we applied two strategies for obtaining seeds. First, we used the citation links of randomly selected papers to induce seed subgraphs. Second, disjunct clusters constructed by applying two different hard clustering algorithms were used as seeds (see below, \ref{sec:2010} and \ref{sec:direct-cites}). Each seed was first adapted by a node-wise local search and then used to initialise the population of different communities by mutating the adapted seed. 

\paragraph{Running the memetic algorithm:} 
Algorithm \ref{pseudcode-mem-evol} is run for each population. Further details and parameter setting can be found in Appendix \ref{app:running}. 

\paragraph{Termination:} 
Due to the many local minima in the cost landscape and the random elements in our procedure, there is always the possibility that more and better valid communities exist. We therefore applied a pragmatic rule by terminating experiments when we did not find new valid communities that increase the coverage of the network 
(cf.\ Algorithm \ref{pseudcode-exp}).

\paragraph{Final selection:} In our approach, a good community is a connected link set with a low value of cost function~$\Psi$.
Displaying coverage over $\Psi$ can be used for selecting the final set of best communities (cf.\ Fig.\ \ref{Fig-rel-cov} for an example).
We calculate the coverage of the network by ranking communities according to $\Psi(L)$ and determining the percentage of all links in the network which are in the union of link sets $L$ up to the current $\Psi$-value. We use the lowest $\Psi$-value above which the coverage does not substantially increase as a cutoff point and disregard all valid communities with higher $\Psi$-values because they represent 
more weakly delineated topics. A specific investigation of thematic structures might require a different decision.

\section{Data}

We downloaded all articles, letters and proceedings papers from the Web of Science published 2003--2010 in journals listed in the category Astronomy \& Astrophysics of the Journal Citation Reports of these years. 
Reference data had to be standardised with rule-based scripts.  The algorithm was applied to two citation networks obtained from this dataset.

\textit{First}, we analysed the  citation network of 14,954 papers published 2010 and the sources cited in these papers. To reduce the complexity of the network, we omitted all sources that are cited only once because they do not link papers and their removal should not unduly influence clustering. We excluded 184 papers that are not linked to the giant component of the citation network and proceeded with a network of 119,954 nodes (14,770 papers and their cited sources) that are connected by 536,020 citation links.  The network would be bipartite 
except for several direct citations between papers published 2010. 
The solution we found in this network is referred to as clustering~\tens{h}. 

\textit{Second}, we analysed  the network of all Astronomy \& Astrophysics papers published 2003--2010 linked by direct citations. 
The giant component of this citation network contains 101,831 papers as nodes and 
924,750 citation links between them. This network is the data model used 
by \shortciteN{Velden2016Infomap} and by  \shortciteN{vanEck2016citation} in their studies published in this issue. 
The solution we found in this network is referred to as clustering \tens{hd}. 

For both networks, we neglected the direction of citation links and analysed an undirected unweighted connected graph. 
Neglecting the direction (and thus the time dimension) of citation directions is unproblematic if communities of thematically similar links are constructed because the theme of a link is determined jointly by the citing and the cited paper. Other approaches might have to consider the direction of citation links in the 2003--2010 direct-citation data-model because the nature of citation links shifts within the network. 2003 papers are almost exclusively linked to the network by being cited. For papers of subsequent years, the nature of links changes until the 2010 papers are linked almost exclusively by citing other papers.

\section{Experiments and Results}
\label{sec:results}

\subsection{The citation network of papers published in 2010 (clustering \tens{h})}
\label{sec:2010}

\begin{figure}[!p] 
\begin{center}  \includegraphics[width=4.5in]{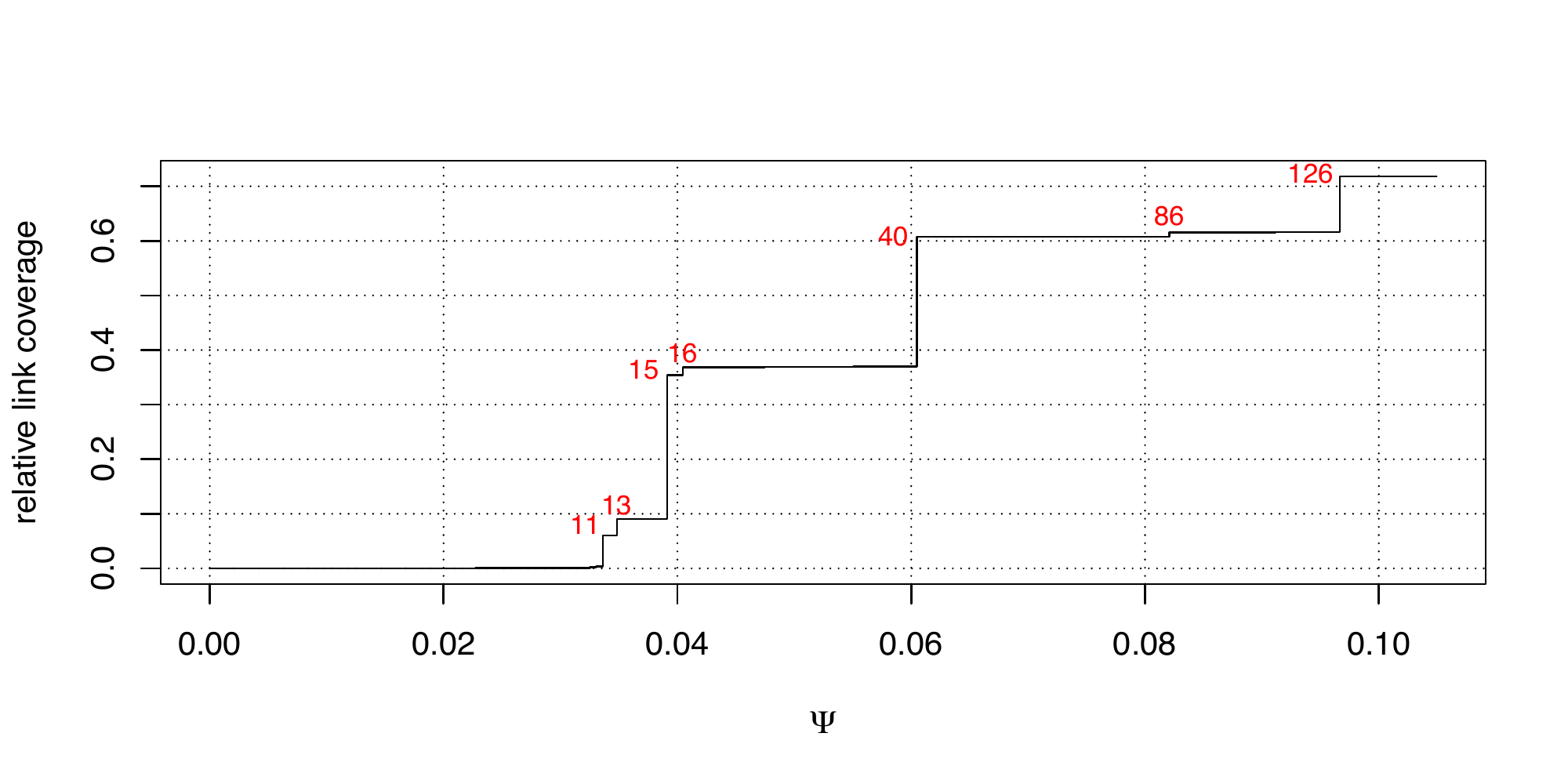} 
\end{center} 
\caption{Relative coverage of network 2010 by  best valid communities (clustering \tens{h} without \tens{h1}, the largest community; the numbers of communities which cover the portions given at the $y$-axis are printed in red)} \label{Fig-rel-cov} 
\end{figure}%

\begin{figure}[!p] 
\begin{center}  \includegraphics[width=4in]{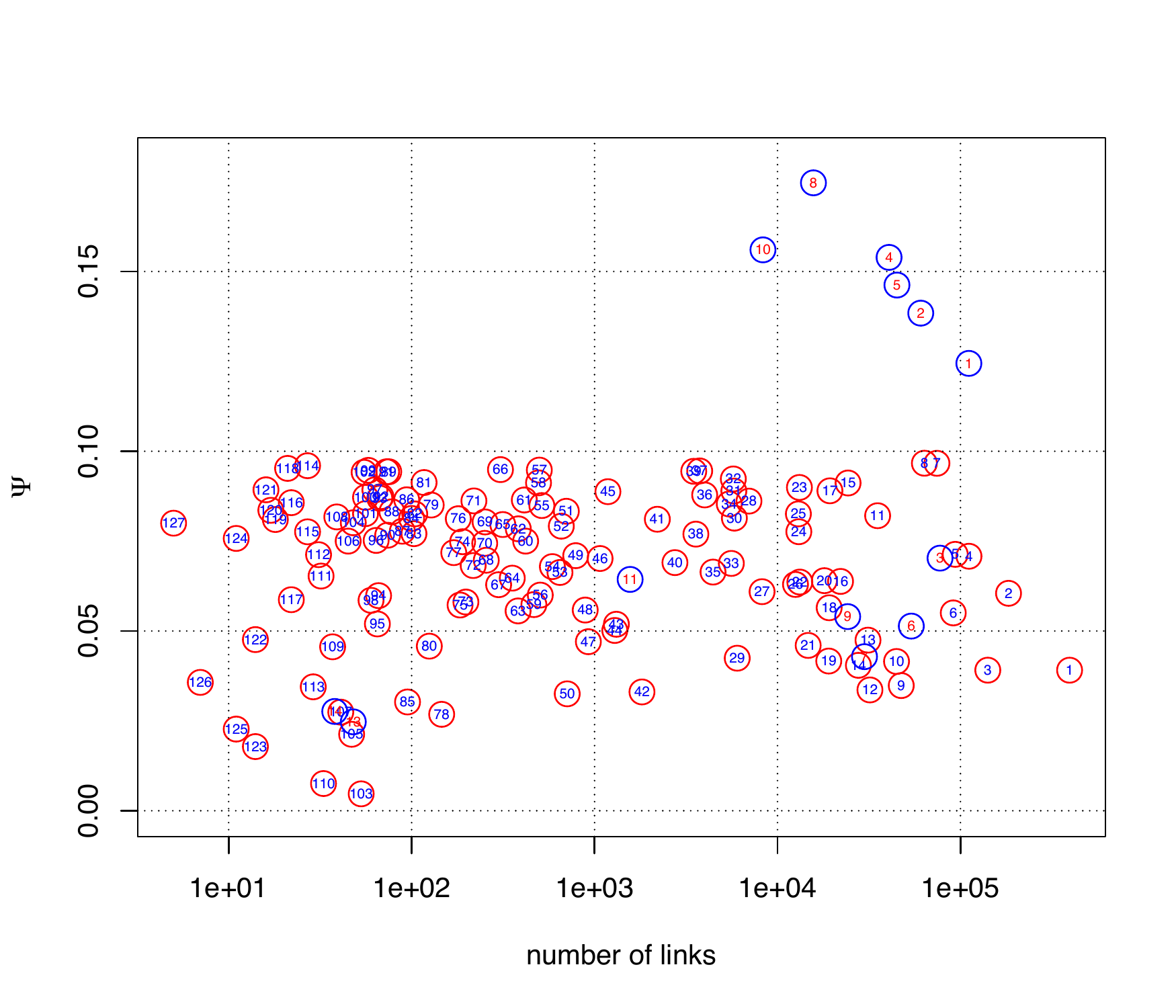} 
\end{center} 
\caption{Cost $\Psi(L)$ over size $|L|$ (in logarithmic scale) for the  127 \tens{h}-communities (red circles) and for 13 of 14 disjoint \tens{uh}-communities obtained with Infomap by Theresa Velden (blue circles, outlier \tens{uh12} with 26 links and $\Psi=0.55$ omitted). Communities are numbered with their size ranks, size measured with number of links $|L|$ for our 127 communities and with number of nodes for 
\tens{uh}-communities. } \label{Fig-cost-size} 
\end{figure}%

\begin{figure}[!t] 
\begin{center}  
\includegraphics[width=2.3in]{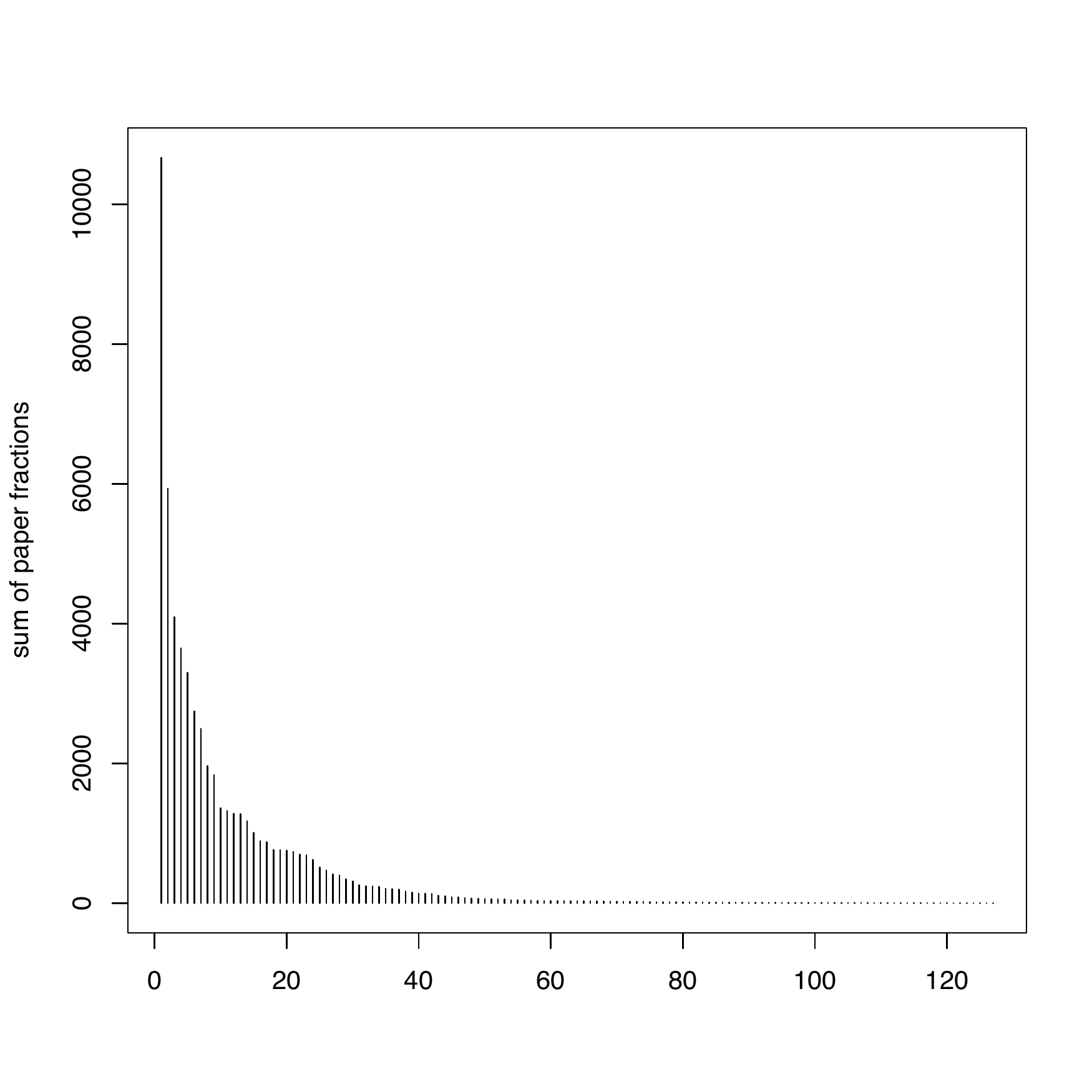}
\includegraphics[width=2.3in]{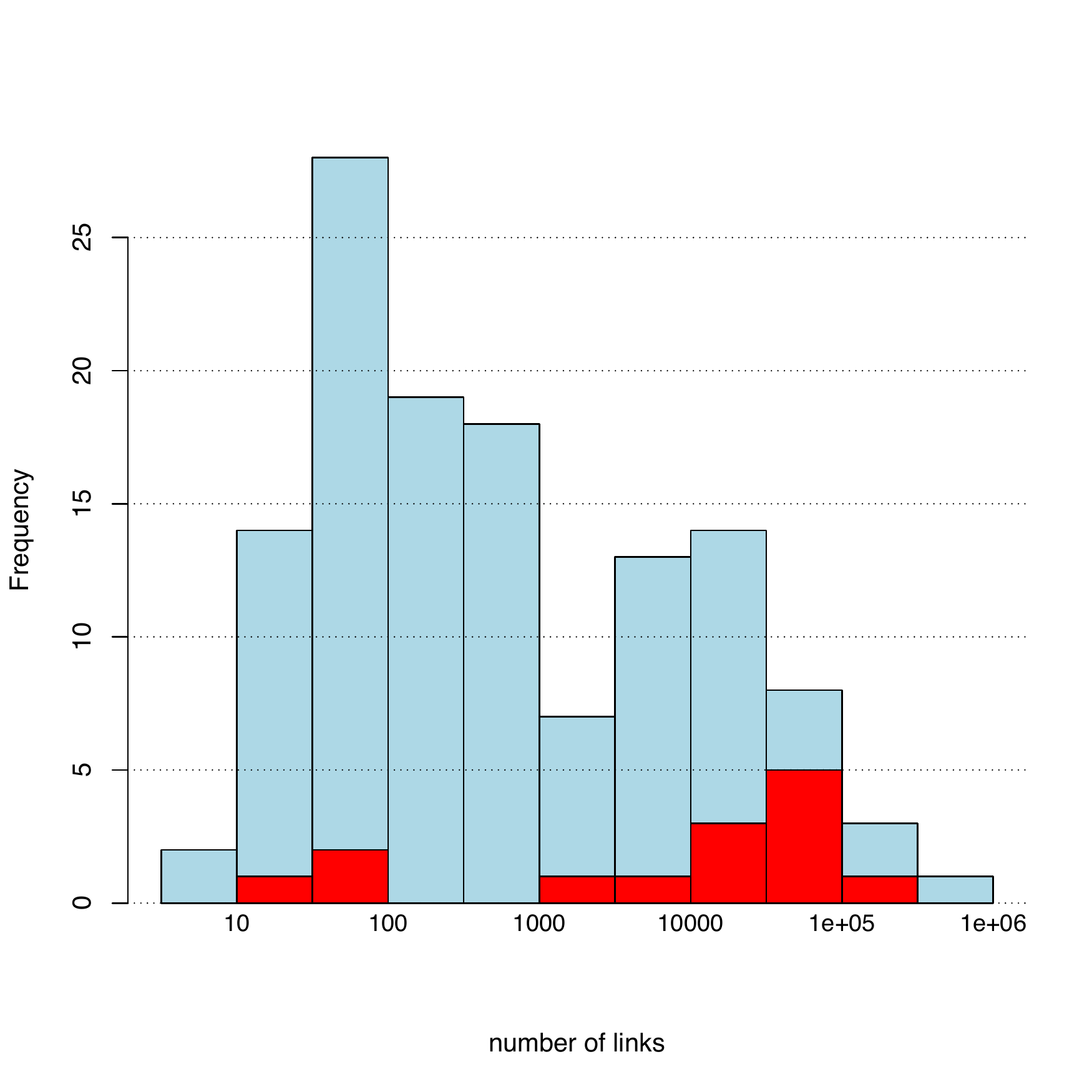} 
\end{center} 
\caption{Size distribution of 127 selected \tens{h}-communities 2010; left side: rank-size plot (measured by sum of paper fractions), right side: histogram based on logarithms of link numbers (histogram of 14 \tens{uh}-clusters shown as red columns)} \label{Fig-sizes-A} 
\end{figure}%

Seed graphs were obtained from Ward clustering  the network's nodes with a similarity measure derived from theoretical considerations \cite{Glaeser2015ISSI}.  
We ordered all Ward clusters by their stability (the length of their branch in the dendrogram) and selected 63 of the most stable clusters as seeds. In addition, we used the citation links of 969 randomly selected papers to induce small seed subgraphs. 

Searches starting from these seeds resulted in a total of more than 
2,600 valid communities 
including many small communities with relatively bad cost values ($\Psi > 0.2 $). Displaying link coverage over $\Psi$ for the union of the best communities 
ranked with regard to $\Psi$ according to Algorithm \ref{pseudcode-exp} we 
observe a sharp knee at 126 communities, with  
coverage equal to 71.8\,\% and cost below 0.0967 (Fig.\ \ref{Fig-rel-cov}). For this calculation, we have omitted the largest community with more than half of all links because including it would lead to a coverage of 100\,\% with only 15 communities, i.e.\ exclude most of the substructures we are interested in.

The largest community \tens{h1} (the number is the size rank, size in terms of numbers of links) contains 394,924 links or 74\,\% of all links. It is the only one that is larger than  half of the network. It contains 
10,840  papers, 10,354 of which are full members with all their citation links  in link set $L$. 

Fig.\ \ref{Fig-cost-size} displays cost $\Psi$  over size (number of links  $=|L|$) for all 127 selected communities  and the clusters obtained by Theresa Velden, who applied the Infomap algorithm \shortcite[this issue]{Velden2016Infomap} to our network. 
Fig.\ \ref{Fig-sizes-A} displays size measured by sums of paper fractions and by link numbers for all 127 communities.  
Table \ref{tab:2010} (s.\ Appendix, p.\ \pageref{tab:2010}) lists number of links, cost, fractional score, and other characteristic data for the 50 largest communities.\footnote{See webpage \url{http://researchdata.ibi.hu-berlin.de/comparison2010.htm} for cluster description sheets. The whole clustering solution and more statistical data can be downloaded from \url{http://researchdata.ibi.hu-berlin.de/Astronomy&Astrophysics/2010/}.}

\subsection{The  network of direct citations  2003--2010 (clustering \tens{hd})}
\label{sec:direct-cites}
We used clusters of papers obtained by 
\citeN[this issue]{vanEck2016citation} 
as seeds.
We selected 469 clusters from three clusterings of this network at different levels of resolution. The smallest cluster contains 50 papers, the largest  20,209 papers.

\begin{figure}[!t] 
\begin{center}  \includegraphics[width=4.5in, height=2.2in]{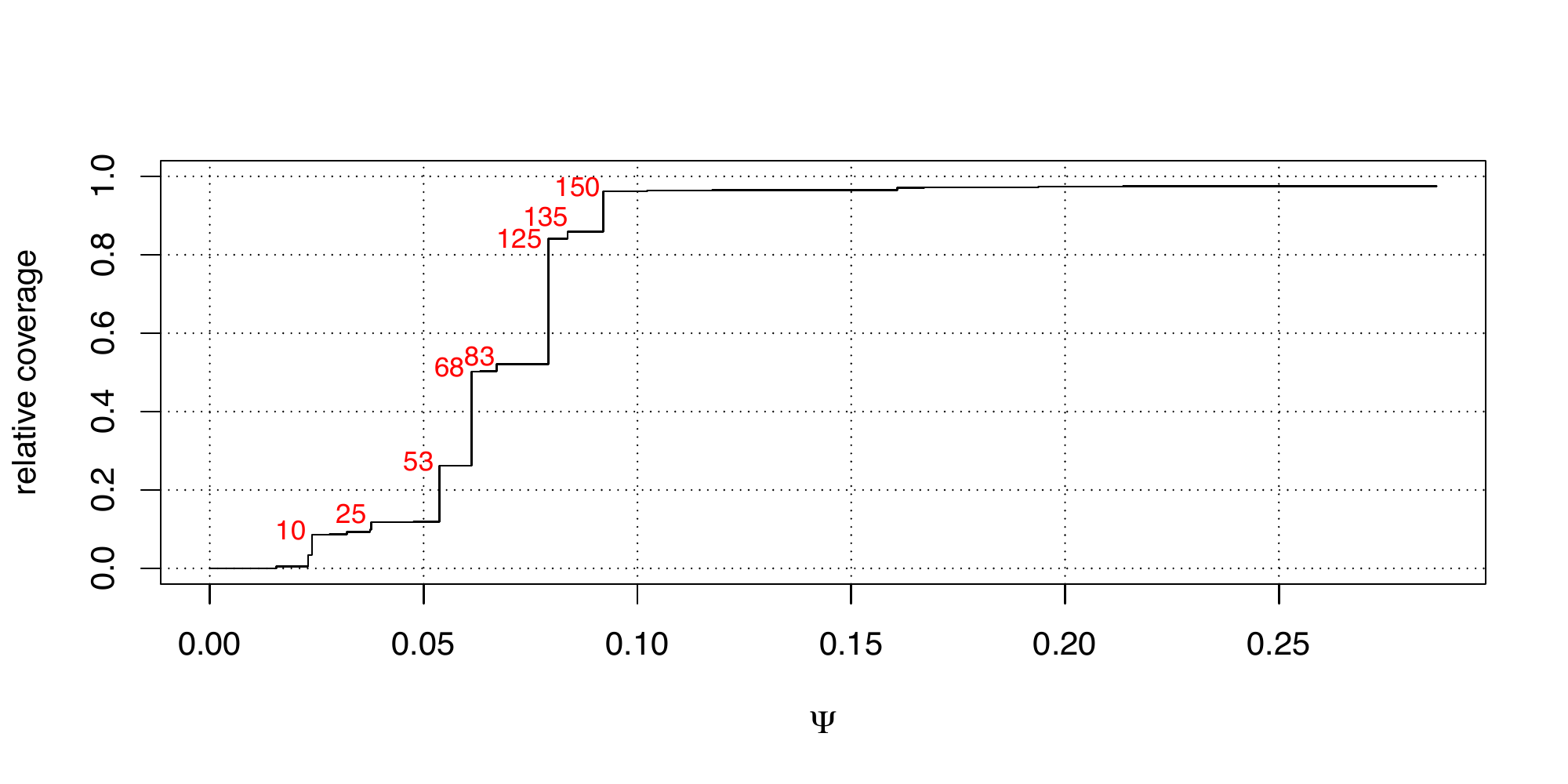} 
\end{center} 
\caption{Relative coverage of network 2003--2010 by 379 valid \tens{hd}-communities (without \tens{hd1} and \tens{hd2}; the numbers of communities which cover the portions given at the $y$-axis are printed in red)} \label{Fig-rel-cov-DiCitNet} 
\end{figure}%

\begin{figure}[!p] 
\begin{center}  \includegraphics[width=4.7in]{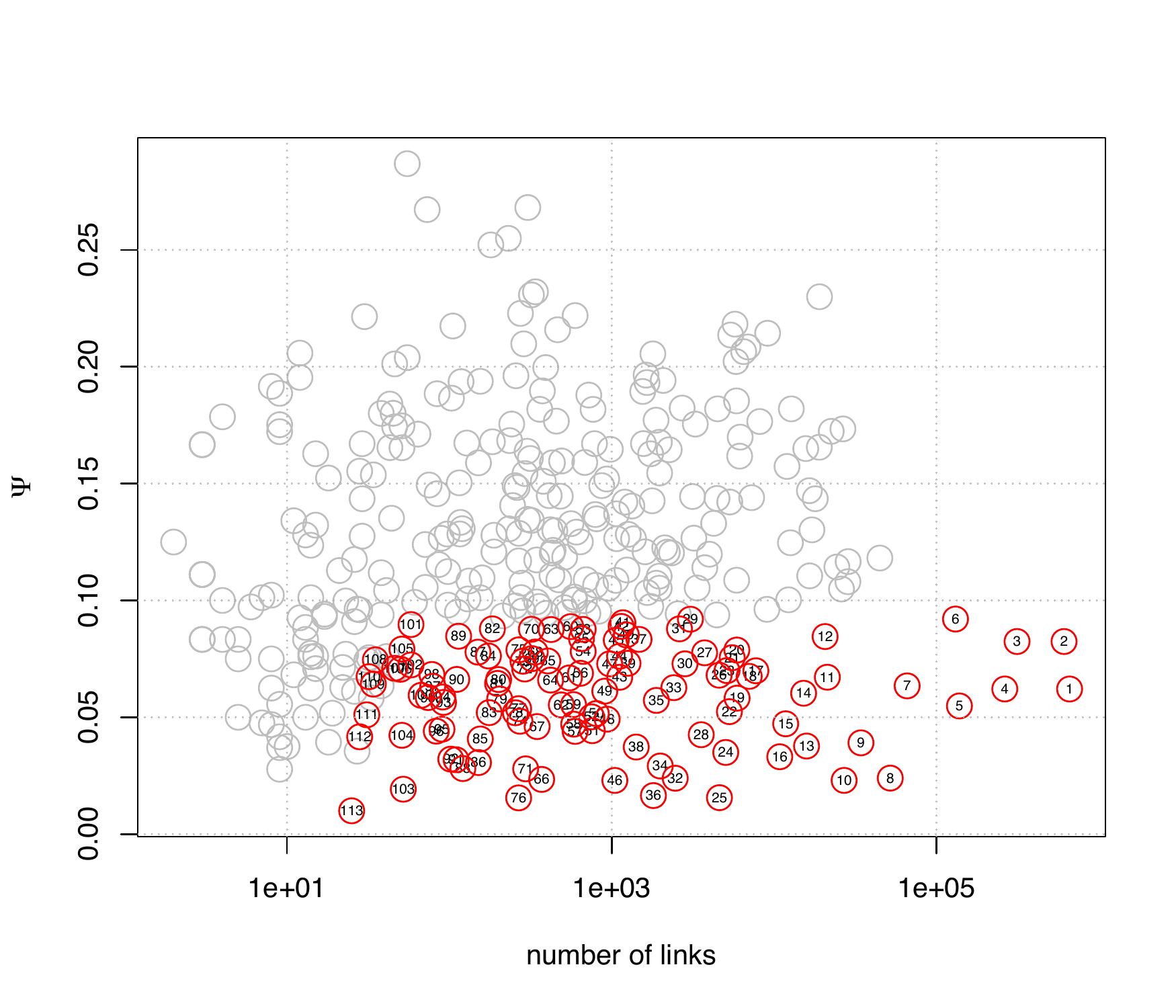} 
\end{center} 
\caption{Cost-size diagram of 381 valid communities. The 113 selected \tens{hd}-communities are represented by red circles \tens{hd}-numbers (size ranks, cf.\ text).} \label{Fig-cost-size-DiCitNet} 
\begin{center}  
\includegraphics[width=2.3in]{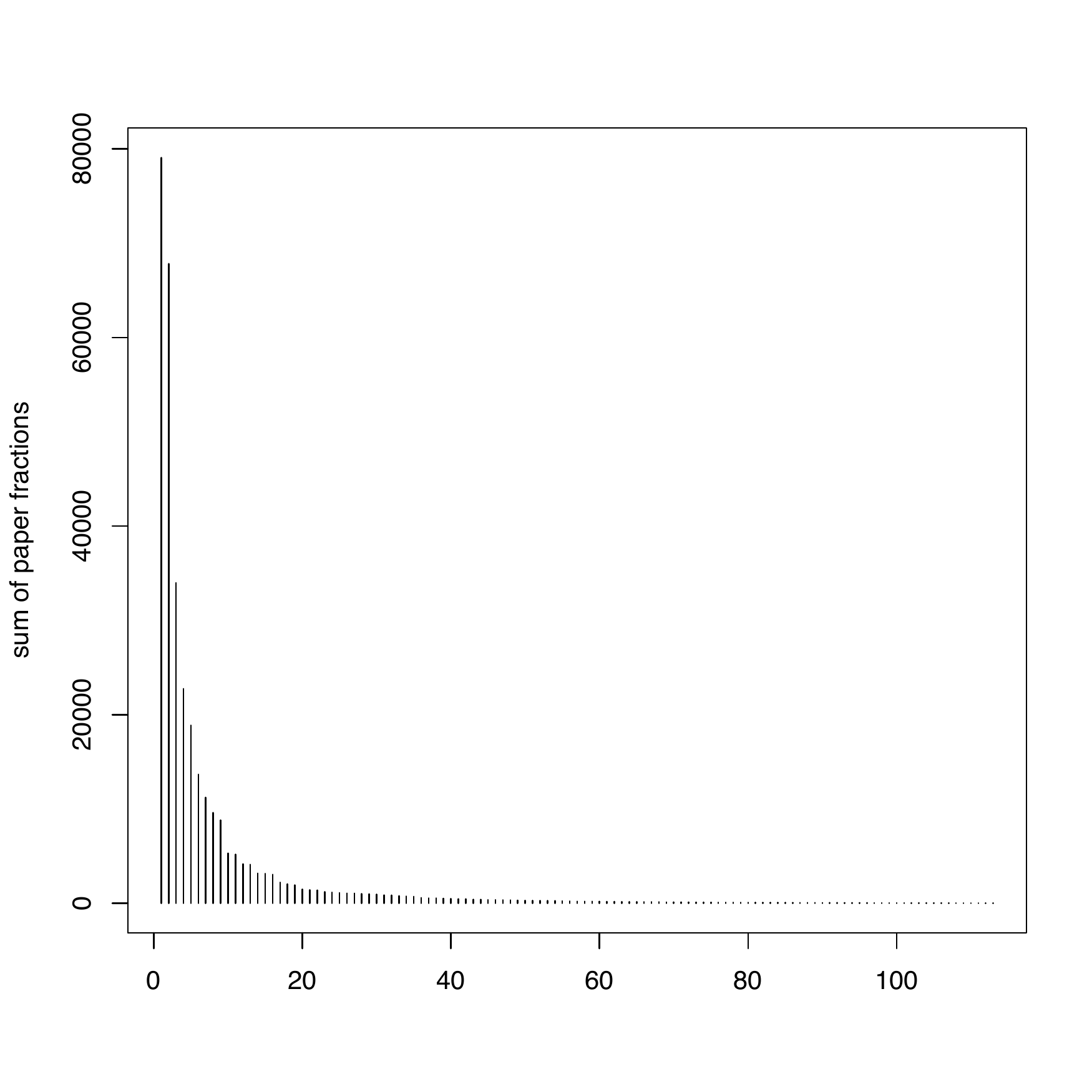}
\includegraphics[width=2.3in]{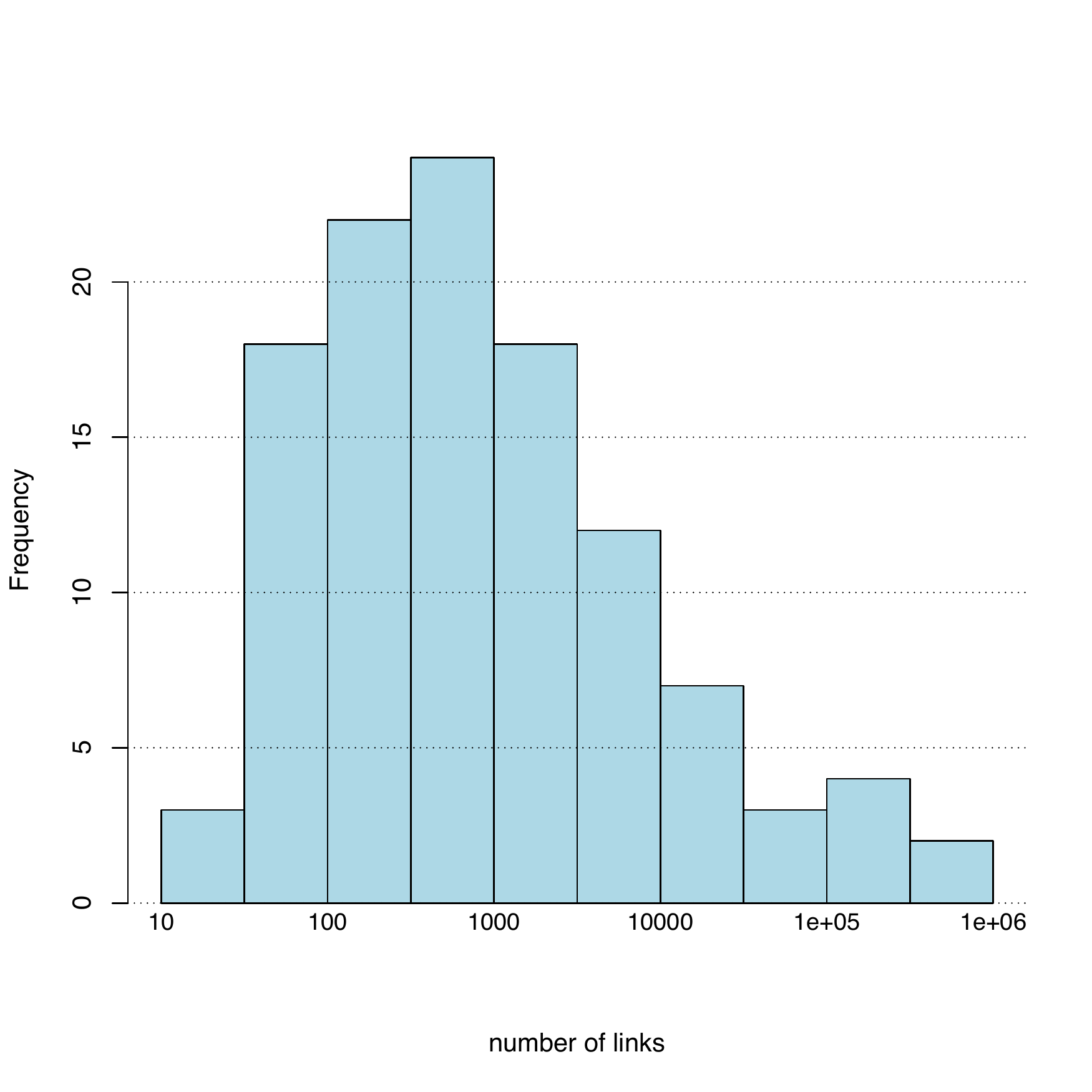} 
\end{center} 
\caption{Size distribution of 113 selected \tens{hd}-communities 2003--2010; left side: rank-size plot (measured by sum of paper fractions), right side: histogram based on logarithms of link numbers} \label{Fig-sizes} 
\end{figure}%

\begin{figure}[!t] 
\begin{center}  
\includegraphics[width=2.3in]{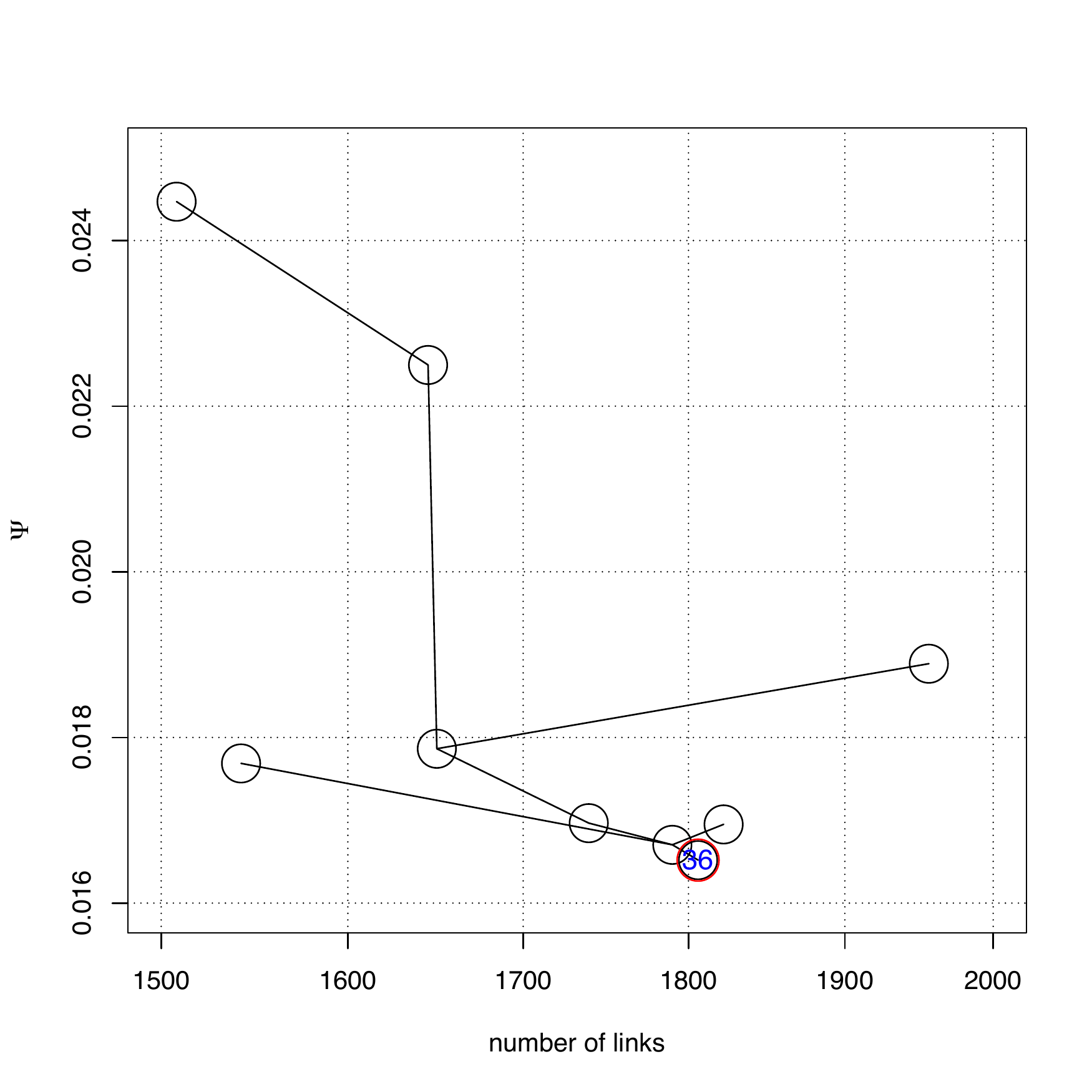}
\includegraphics[width=2.3in]{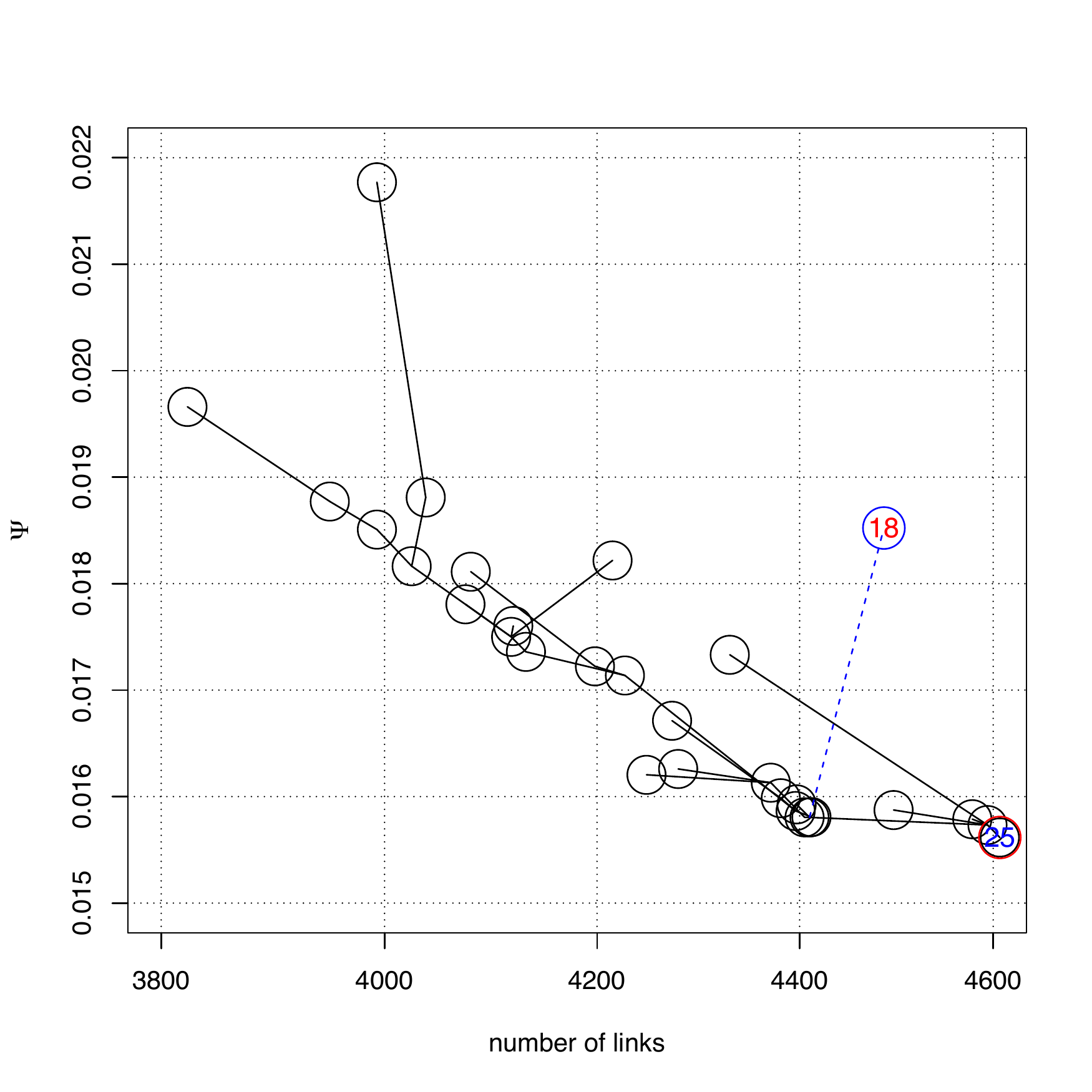}
\end{center}
\caption{Two details of cost-size diagram 
around valid communities \tens{hd36} and \tens{hd25} (red circles). Each invalid community (black circle) is connected by a  black line to the nearest (most similar) community with lower $\Psi$-value that makes it invalid. 
The blue circle represents cluster \tens{c18}, which  we have used  as a seed.
The improvement of its $\Psi$-value by our algorithm is marked by the blue dashed line.} 
\label{Fig-selection-cost-size-DiCitNet} 
\end{figure}
 
For some seeds the algorithm had to be terminated prematurely because there were not enough mutants to initialize a population or because local searches led to subgraphs containing more than three quarters of all links. In these cases, the final link-wise search was applied to results of intermediate steps. 
We found 381 valid communities.

There are two valid communities with more than half of all links (\tens{hd1} and \tens{hd2},  the number is the size rank, size measured with numbers of links). They are   the main components of the link complements of \tens{hd4} and \tens{hd3}, respectively. 
To select the best communities with maximum coverage we rank all but the two largest valid communities according to their $\Psi$-values 
and successively unite them reaching a coverage of 97.5\,\% of all links 
(Fig.\ \ref{Fig-rel-cov-DiCitNet}). 
Of the 379 remaining communities we selected those  with $\Psi < 0.0921$  because communities above this threshold improve relative coverage only by 1.3\,\%. Again, including the two largest communities would lead to the exclusion of interesting sub-structures. 

The large number of very small communities in our sample required a decision on the minimum size from which a community should be considered a topic. We set this threshold at 20 (fractionally counted) papers over 8 years, which led to the exclusion of further 39 communities (cf.\ cost-size diagram in Fig.\ \ref{Fig-cost-size-DiCitNet}).\footnote{Coverage is decreased by this omission only by a negligible amount.}  In Fig.\ \ref{Fig-sizes} size measured by sums of paper fractions and by link numbers is displayed for all 113 selected communities (see Fig.\ \ref{Fig-sizes-A} for the distribution of community sizes  of the 2010 network). 
A description of the 50 largest communities can be found in Table \ref{tab:8y} in Appendix, p.\ \pageref{tab:8y}.\footnote{See webpage \url{http://researchdata.ibi.hu-berlin.de/comparison.htm} for cluster description sheets. The whole clustering solution and more statistical data can be downloaded from \url{http://researchdata.ibi.hu-berlin.de/Astronomy&Astrophysics/2003-2010/}. \label{sheets}}

For several seeds memetic searches ended in the same sink of the cost  landscape. The most attractive sink is that around community \tens{hd25} where 29 search paths ended (s.\ Fig.\ \ref{Fig-selection-cost-size-DiCitNet}). 
One of the seed subgraphs---cluster \tens{c18} of clustering solution \tens{c} \cite[this issue]{vanEck2016citation}---is already located in this sink but was improved by memetic search. The paper-based Salton index of the pair \tens{hd25}--\tens{c18} is 0.94.\footnote{The Salton index measures similarity of two sets and is defined as the ratio of the size of their intersection and the geometric mean of their sizes. It is also called Salton's cosine because it can be calculated as the cosine of the angle between two vectors characterising the sets.\label{Salton}} Their papers are about phenomena in the magnetosphere.\footnote{cf.\ webpage \url{http://researchdata.ibi.hu-berlin.de/comparison.htm} with clusters \tens{c18} and \tens{hd25}}

\section{Exploring the Cluster Solutions}
\label{sec:expl}
In this section, we explore our two cluster solutions by discussing their correspondence 
(section \ref{sec:consistence}) and by analysing the poly-hierarchies of both solutions (section \ref{sec:poly}). We debate possible reasons for unstructured regions  (section \ref{sec:ti}) and compare the 2010 solution with a clustering of the same network obtained with another method (section \ref{sec:Theresa2010}). 

As mentioned above, we refer to the 127 communities found in the citation network of 14,770 papers published 2010 in astrophysics journal as clustering \tens{h} and name the solution found in the network of 101,831 astrophysics papers 2003--2010 linked by direct citations clustering \tens{hd}. The two networks share most of papers published 2010 and all papers  cited in 2010 papers and published 2003--2009 in journals of the set considered. The network of direct citations between papers published 2003--2010 does not contain cited sources outside this paper set. It is more than two times denser than the 2010 network.

\begin{table}[!t]
\caption{The fourteen \tens{h}--\tens{hd} pairs of communities with a Salton index above 0.8.  
The overlap numbers in the last two columns are size of intersection of sets of 2010 papers divided by size of community in \tens{h} and \tens{hd}, respectively (cf.\ text). The Salton index is the geometric mean of the last two columns.}
\label{Tab:match} 
\begin{tabular}{rrrrrrrr}
\hline\noalign{\smallskip}
size(\tens{h})    &	  & 	       &   & size(\tens{hd})& 	Salton&	overlap&overlap \\ 
rank&	 \tens{h}&	size(\tens{h})& \tens{hd}& in 2010&	index& in \tens{h}&in \tens{hd}\\
\noalign{\smallskip}\hline\noalign{\smallskip}
1 &  2 &5932 & 4  &4677  & 0.87     & 0.77      & 0.98	\\
2 &  4 &3583 & 4  &4677  & 0.81     & 0.92      & 0.70 	\\
3 &  6 &2737 & 5  &2599  & 0.84     & 0.82       &0.86	\\
4 & 12 &1352 & 8  &1144  & 0.88     & 0.81      & 0.96	\\
5 & 14 &1275 & 9  &1158  & 0.85     & 0.81      & 0.89	\\
6 & 19 & 868 &13  & 711  & 0.84     & 0.76      & 0.93	\\
7 & 18 & 609 &10  & 700  & 0.84     & 0.90       &0.78	\\
8 & 29 & 243 &28  & 218  & 0.85     & 0.81       &0.90	\\
9 & 34 & 206 &17  & 210  & 0.87     & 0.87       &0.86	\\
10& 38 & 132 &23  & 132  & 0.90      &0.90      & 0.90	\\
11& 50 &  33 &86  &  22  & 0.82     & 0.67      & 1.00	\\
12 &66 &  30 &54  &  24   &0.82     & 0.73      & 0.92	\\
13 &61 &  15& 61  &  14  & 0.90     & 0.87      & 0.93	\\
14 &92 &   4 &91  &   4  & 1.00     & 1.00      & 1.00	\\

\noalign{\smallskip}\hline
\end{tabular}
\end{table}

\subsection{Correspondence of the two solutions}
\label{sec:consistence}

Despite the differences between the two networks, our two clustering solutions \tens{h} and \tens{hd} have many \tens{h}--\tens{hd} pairs of communities with a good match. 
We compared individual communities from both solutions by assigning as community members all papers that were published in 2010 and have membership grades larger than 0.5 in a community of clustering \tens{h} or \tens{hd}, respectively. Among these,
14 \tens{h}--\tens{hd} pairs of communities 
have a Salton index of 0.8 or higher (cf.\ Table \ref{Tab:match}). 
Among these 14  best matches 
there  are pairs of very large communities. 11 of 14 best matching pairs show an overlap of at least 0.9, that means that at least 90\,\% of the smaller is covered by the larger community.

Due to the differences between the networks we cannot expect that all 113 \tens{hd}-communities match one of the 127 \tens{h}-communities with high similarity. Some \tens{hd}-communities have very few papers in 2010 (we have, e.g.,  25 of them with less than five papers in 2010 and two without any 2010 paper). Others are not connected if we only consider their papers in 2010.\footnote{We have put papers of each \tens{hd}-community into the 2010-network and linked them via all their cited sources and calculated the sizes of their main components.}  
To get an impression of matching, we have plotted similarity of \tens{hd}-communities to best matched \tens{h}-communities over size of \tens{hd}-communities. 
Bad matching is mainly occurring for small \tens{hd}-communities which typically have large \tens{h}-partners.

\subsection{Overlaps and poly-hierarchies}

\label{sec:poly}
\begin{figure}[!t] 
\begin{center}  \includegraphics[width=3in]{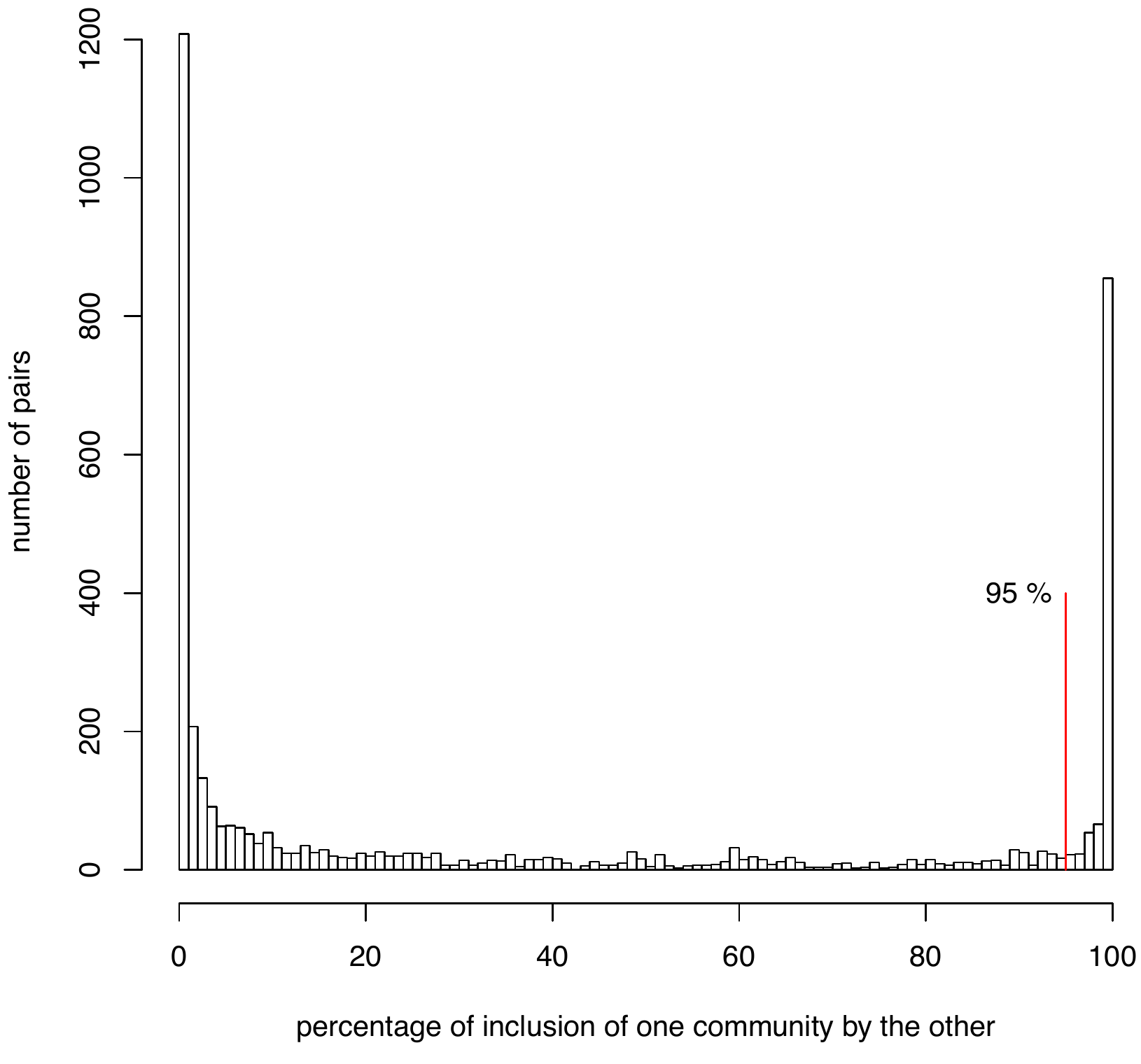} 
\end{center} 
\caption{Histogram of relative inclusion of pairs of overlapping communities in clustering \tens{h}} \label{Fig-U} 
\end{figure}

In both solutions there are many pairs of communities that overlap i.e.\ share links and nodes. Overlaps vary from very small to inclusion, i.e.\ communities being subgraphs of larger communities. In Fig.\ \ref{Fig-U} we show the distribution of relative inclusion for pairs of overlapping communities in clustering \tens{h}. Relative inclusion is the percentage of links of one community which are also in the other community. The density function has a typical U-form with many pairs with small inclusion and many pairs with total or nearly total inclusion. 
\label{poly} 
Analysing hierarchical relations between link communities by comparing their link sets, we found that large communities have many subcommunities. All these subcommunities are small and are typically two orders of magnitude smaller than the large community. This picture changes substantially if we relax the subset relation and only demand that a subtopic community should share at least 95\,\% of its links with each of its super\-topic communities. The threshold is derived from the distribution in Fig.\ \ref{Fig-U}. At 95\,\% the number of pairs begin to rise. 
In Tables \ref{tab:2010} and \ref{tab:8y} (Appendix \ref{appendix}) 
we provide the numbers of subtopics, of super\-topics, and of other communities which overlap for each community in the two solutions.

\begin{figure}[!t] 
\begin{center}  \includegraphics[height=1.8in]{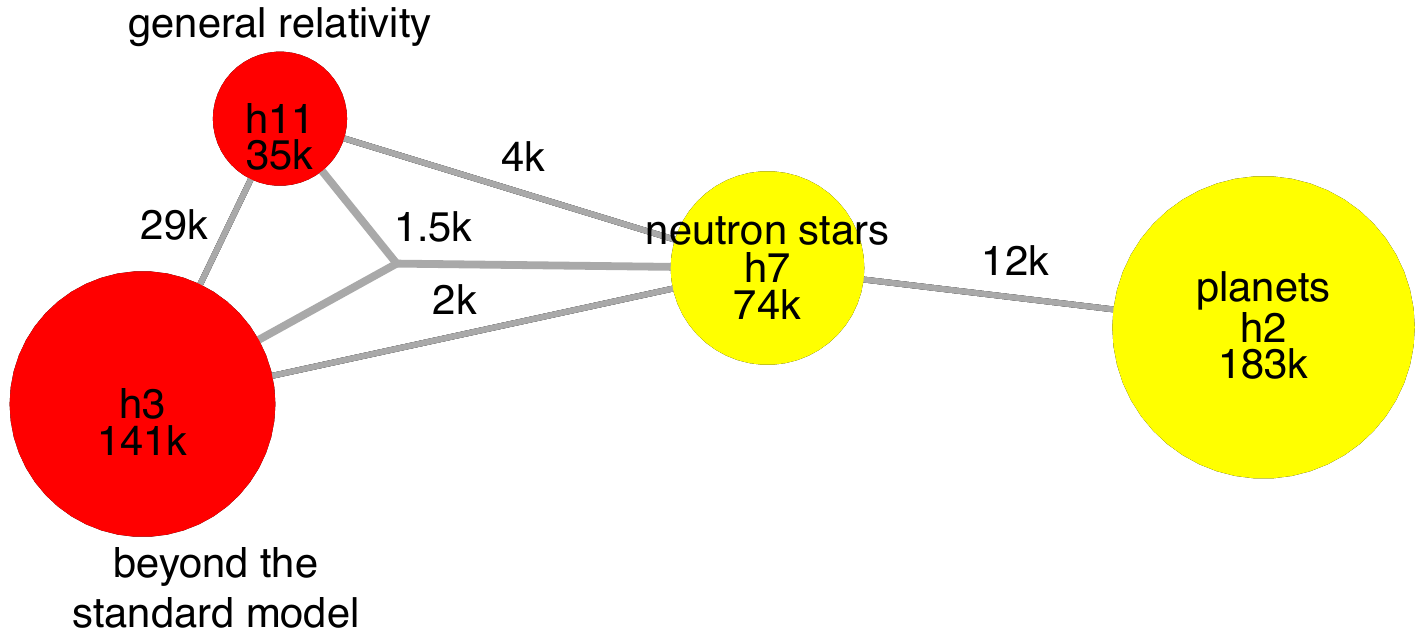} 
\end{center} 
\caption{0verlaps between large communities  of clustering \tens{h}: the three communities \tens{h3}, \tens{h7}, and \tens{h11} share 1511 links (1.5k, visualised as a hyperedge, the sizes of their pairwise overlaps are diminished by 1511). Communities are connected by an edge or a hyperedge if they share more than 100 links. Communities \tens{h2} and \tens{h7} (yellow) are subtopics of \tens{h1} (not shown). All other but four small communities are subtopics of the four communities shown (some are subtopics of more than one large community, cf.\ text and Table \ref{tab:2010}). 
For the labels  cf.\  bold terms in Table~\ref{Tab:terms-h}. 
Graph layout: Fruchterman-Reingold algorithm (R-Package \textit{sna}).} \label{Fig-h-overlap} 
\end{figure}%
\begin{table}[!t]
\caption{High-ranked terms describing the themes of four \tens{h}-communities shown in Fig.\ \ref{Fig-h-overlap}. Terms have been extracted from Unified Astronomy Thesaurus  (before semicolon) and from titles and abstracts (after semicolon), cf.\ text.} 
\label{Tab:terms-h} 
\begin{tabular}{rp{10.7cm}}
\hline\noalign{\smallskip}
\tens{h}& keywords \\
\noalign{\smallskip}\hline\noalign{\smallskip}
2& solar system, \textbf{planets}, solar physics, stellar structure, circumstellar matter, planetary atmospheres, extrasolar planets, asteroseismology, mars, solar wind; solar, planet, stars, sun, atmosphere, mars, period, earth, spacecraft, coronal\\
3&\textbf{beyond the standard model}, cosmic microwave background radiation, relativity, large scale structure of the universe, p branes, dark energy, general theory of relativity, quantum gravity, cosmological models, neutrino masses; scalar, scalar field, standard model, qcd, lhc, higgs, inflation, gravity, quark, quantum
\\
7&\textbf{neutron stars}, astroparticle physics, gamma ray bursts, mars, pulsars, stellar phenomena, blazars, gravitational waves, x ray binary stars, binary systems; gamma ray, neutron star, mars, ray bursts, pulsar, grb, x ray, gravitational wave, high energy, swift
\\
11&\textbf{relativity}, black holes, quantum gravity, beyond the standard model, p branes, gravitation, \textbf{general} theory of relativity, gravitational singularities, event horizons, hawking radiation; spacetime, black hole, quantum gravity, horizon, metric, gravity, solutions, schwarzschild, loop quantum, holographic
\\
\noalign{\smallskip}\hline
\end{tabular}
\end{table}

\begin{figure}[!p] 
\begin{center}  \includegraphics[height=3.7in]{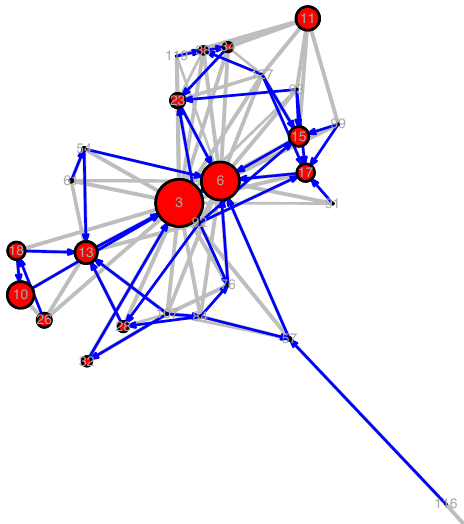} 
\end{center} 
\caption{Graph of poly-hierarchy  in clustering \tens{h} with subtopics of \tens{h3} and \tens{h11}. The numbers are \tens{h}-identifiers (cf.\ Table \ref{tab:2010} in Appendix \ref{appendix}), blue and grey links indicate an at least  95\,\% inclusion  of the smaller community in the larger one. Blue arrows visualise the direct subtopic-super\-topic relations in the poly-hierarchy of subtopics of community~\tens{h3}. 
 Community~\tens{h116} (in the bottom right-hand corner) is also a subtopic of \tens{h2}.} \label{Fig-hierarchy-of-3}

\begin{center}  \includegraphics[width=2.9in]{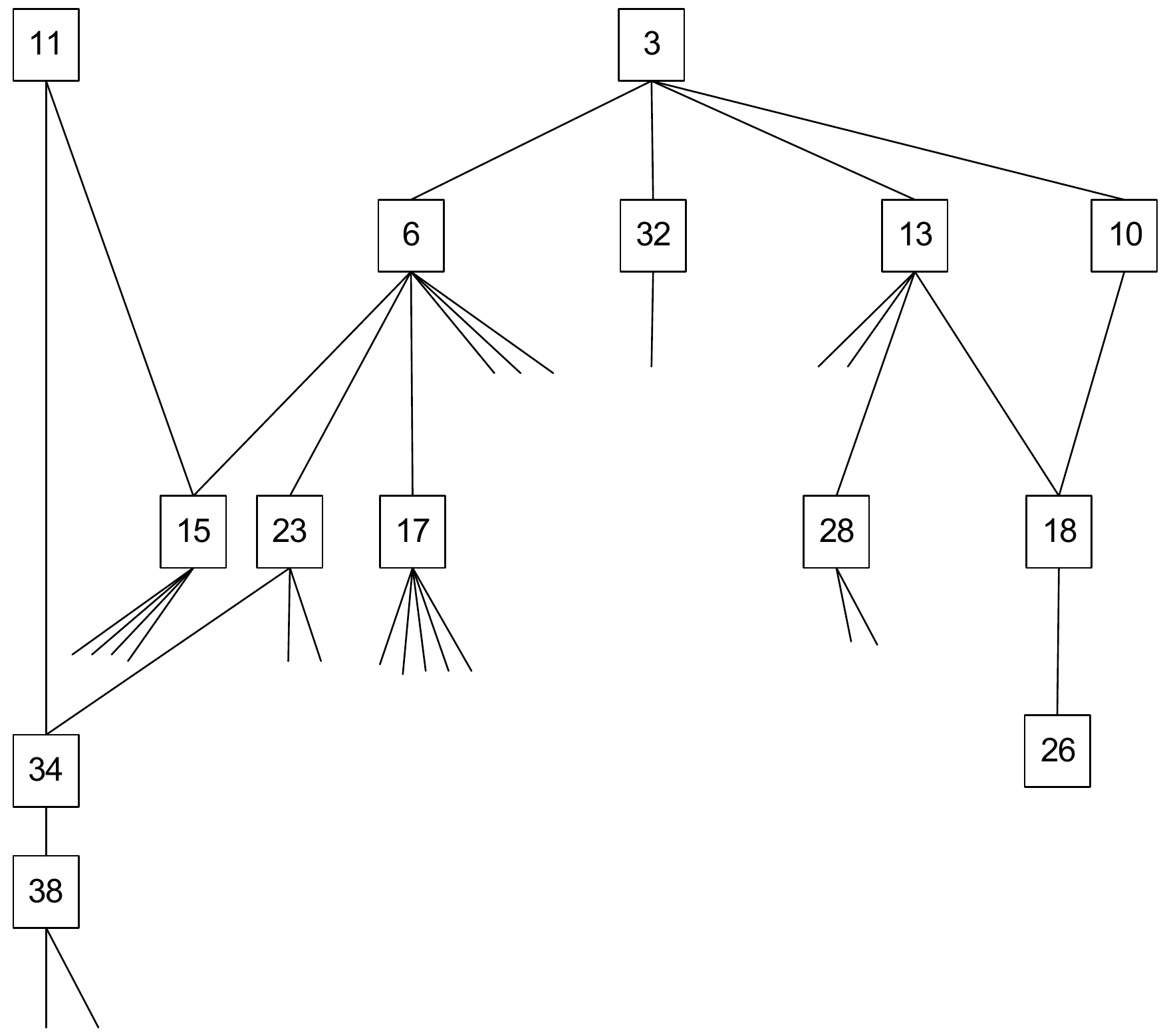} 
\end{center} 
\caption{Organisational chart of poly-hierarchy of the 14 largest \tens{h}-communities in Fig.\ \ref{Fig-hierarchy-of-3}} \label{Fig-hierarchy-of-3-17} 
\end{figure}

Communities  at the highest levels of their poly-hierarchy and  their overlap relationships are of interest when thematic structures within the paper sets are analysed.
Solution \tens{h} has three communities without super\-topic: \tens{h1}, \tens{h3}, and \tens{h11}. 
Because large communities obscure relationships of subtopics we want to analyse, we excluded communities that are larger than half of the networks (\tens{h1}, \tens{hd1}, and \tens{hd2}) from the following analysis.
Omitting \tens{h1}, there are eight communities at the highest poly-hierarchy level of clustering \tens{h}. They cover 72\,\% of all links. Fig.\ \ref{Fig-h-overlap} visualises the overlaps between the four largest of them and their relation to \tens{h1}.\footnote{There are also four small \tens{h}-communities which have only \tens{h1} as their super\-topic. Two of them have only papers in solid-state physics, the other two are overlapping with \tens{h2} and \tens{h7}.\label{f:isolated}}
In Table \ref{Tab:terms-h} the terms describing the four communities are listed. The terms before semicolons were assigned to papers by Kevin Boyack by applying the Data Harmony's Machine Aided Indexer (M.A.I.) on Unified Astronomy Thesaurus data \shortcite[this issue]{Velden2016Comparison}.\footnote{cf.\ also \url{http://astrothesaurus.org/} and \url{http://www.dataharmony.com/}} The terms after the semicolons are  extracted  by Rob Koopman and Shenghui Wang  from titles and abstracts. They than ranked both types of terms by calculating normalised mutual information \shortcite[this issue]{Koopman2016Mutual}.

We define a \textit{direct} super\-topic of a topic as a super\-topic that cannot be reached indirectly through a  chain of supertopic-subtopic relations. 
In Fig.\ \ref{Fig-hierarchy-of-3} we draw super\-topic-subtopic relations as links and draw blue arrows for direct relations which end in the large community \tens{h3}.  
Community \tens{h18}, e.g., is a subtopic community of communities \tens{h13} and \tens{h10}. The existence of such relationships discriminates poly-hierarchies from normal strict hierarchies. The organisational chart (Fig.\ \ref{Fig-hierarchy-of-3-17}) shows several cases of communities
being included (at the 95\,\% level) in more than one other community,  which corresponds to subtopics with more than one supertopic. 
If we relax the subtopic-supertopic relationship by allowing up to 5\,\% of a subtopic's links not being shared with its supertopic, than the resulting relaxed poly-hierarchy is not always transitive. 
The small community \tens{h116} in the  lower right of Fig.\ \ref{Fig-hierarchy-of-3}, for example, is related to \tens{h3} via \tens{h57} and \tens{h6} (blue arrows) but not directly with a grey link.

\begin{figure}[!b] 
\begin{center}  \includegraphics[height=2.7in]{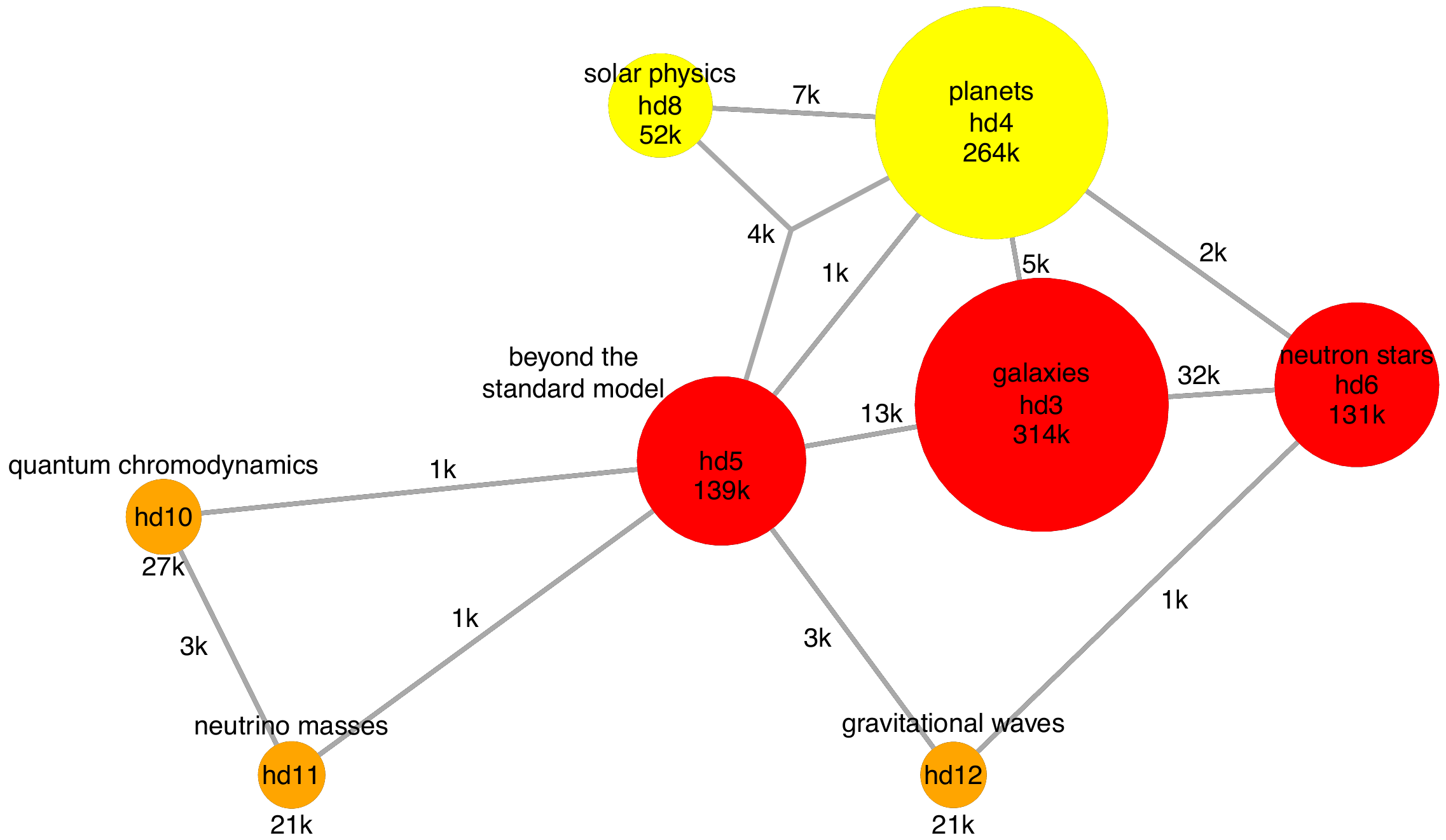} 
\end{center} 
\caption{0verlaps between eight large communities  of clustering \tens{hd}:  the three communities \tens{hd4}, \tens{hd5}, and \tens{hd8} share 4083 links (4k, visualised as a hyperedge, the sizes of their pairwise overlaps are diminished by 4083, which lets the direct overlap edge \tens{hd5-hd8} disappear because only six links in the intersection \tens{hd5-hd8} are not in \tens{hd4}). Communities are connected by an edge or a hyperedge if they share more than 500 links. The red communities are subtopics of \tens{hd1}, the yellow ones of \tens{hd2}, and the orange ones are in the overlap of both, \tens{hd1} and \tens{hd2}. All other but five small communities are subtopics of the eight communities shown (cf. Table \ref{tab:8y} and text). For the labels cf.\ bold terms in Table \ref{Tab:terms-hd} and webpage \url{http://researchdata.ibi.hu-berlin.de/comparison.htm}. Graph layout: Fruchterman-Reingold algorithm (R-Package \textit{sna}).}
\label{Fig-hd-overlap} 
\end{figure}%
\begin{table}[!t]
\caption{High-ranked terms describing the themes of eight \tens{hd}-communities shown in Fig.\ \ref{Fig-hd-overlap}. Terms have been extracted from Unified Astronomy Thesaurus  (before semicolon), titles and abstracts (after semicolon), cf.\ text.} 
\label{Tab:terms-hd} 
\begin{tabular}{rp{10.7cm}}
\hline\noalign{\smallskip}
\tens{hd}& keywords \\
\noalign{\smallskip}\hline\noalign{\smallskip}
3& \textbf{galaxies}, redshift, doppler shift, active galactic nuclei, quasars, galaxy clusters, milky way galaxy, star formation, galaxy groups, astronomical research; galaxies, redshift, active galactic, star formation, agn, galactic nuclei, sample, clusters, quasar, sloan digital
\\
4  &\textbf{planets}, solar system, circumstellar matter, planetary atmospheres, planetary system formation, natural satellites, mars, extrasolar planets, emission line objects, stars; star, planet, main sequence, mars, hd, atmosphere, jupiter, abundances, radial velocity, period\\
5  &\textbf{beyond the standard model}, cosmic microwave background radiation, p branes, dark energy, relativity, cosmological models, radio astronomy, quantum gravity, general theory of relativity, gravitational singularities; scalar field, dark energy, inflation, cosmological constant, gravity, spacetime, microwave background, cosmic microwave, brane, universe\\
6  &\textbf{neutron stars}, gamma ray bursts, pulsars, stellar phenomena, astroparticle physics, blazars, supernova remnants, x ray binary stars, accretion disks, magnetars; gamma ray, x ray, neutron star, ray bursts, grb, pulsar, high energy, jet, swift, bursts grbs
\\
8  &\textbf{solar physics}, stellar structure, coronal mass ejections, starspots, magnetic fields, sunspots, solar flares, asteroseismology, coronae, solar corona; solar, coronal mass, magnetic, active region, cme, sunspot, flare, mass ejections, solar activity, plasma\\
10 &perturbation methods, light cones; \textbf{qcd}, quark, meson, lattice, decays, chiral, pi pi, gluon, pion, j psi
\\
11 &\textbf{neutrino masses}, beyond the standard model, supersymmetric standard model, supersymmetry, leptogenesis, technicolor, grand unified theory, extra dimensions, supernova neutrinos, baryogenesis; standard model, lhc, higgs, lepton, top quark, minimal supersymmetric, neutrino, hadron collider, supersymmetric standard, electroweak
\\
12 &\textbf{gravitational waves}, ligo, interferometers, astroparticle physics, relativity, general theory of relativity, binary stars, black holes, circular orbit, schwarzschild black holes; gravitational wave, lisa, wave detectors, post newtonian, inspiral, ligo, numerical relativity, waveforms, binary black, interferometric gravitational
\\

\noalign{\smallskip}\hline
\end{tabular}
\end{table}

A specific property of the journal \textit{Physical Review D}, in which 18\,\% of the articles in our dataset 2003--2010 are published (19\,\% in 2010), offers an opportunity to assess the performance of our algorithm. Issues of \textit{Phys.\,Rev.\,D} are thematically separated. Only the issues with even numbers publish papers on astronomy and astrophysics, while issues with uneven numbers are devoted to particle physics (1175 papers in issues with uneven numbers in our dataset 2010).\footnote{cf.\ \textit{Subject Areas} on this webpage of \textit{Physical Review D}: \url{http://journals.aps.org/prd/authors/editorial-policies-practices}} 
 Our algorithm reconstructed this distinction, as can be illustrated in Fig.\ \ref{Fig-hierarchy-of-3-17}. The papers in the five clusters \tens{h10}, \tens{h13}, \tens{h18}, \tens{h26}, and \tens{h28} and their subclusters are dominated by particle-physics papers.\footnote{E.g., 1050 of the 1175 particle-physics papers in \textit{Phys.\,Rev.\,D} are among the 1272 full papers in \tens{h10}.} \textit{Quantum chromodynamics} (QCD), \textit{quark}, and \textit{meson} are among the most frequent and distinguishing terms obtained for these topics (for method cf.\ Table \ref{Tab:terms-h}). This branch of the poly-hierarchy reconstructs the thematical separation between issues of \textit{Phys.\,Rev.\,D.}

Fig.\ \ref{Fig-hierarchy-of-3} shows only subtopic-supertopic relationships of 27 communities of clustering \tens{h}. They are subtopics of \tens{h3} and of \tens{h11} and are connected with 97 further communities via \tens{h116}. The three remaining communities are \tens{h1} (which we omitted from the analysis) and two isolated communities mentioned in footnote \ref{f:isolated}.

 Our clustering  \tens{hd} of the direct-citation network 2003--2010 has 13 communities at the highest level of the poly-hierarchy if \tens{hd1} and \tens{hd2} are omitted. The 13 communities  cover 96\,\% of all links. In Fig.\ \ref{Fig-hd-overlap} the overlaps between the eight largest of them are visualised. Table \ref{Tab:terms-hd} lists terms describing the eight topics. They are selected in the same manner as terms in Table \ref{Tab:terms-h} (see above). The corresponding graph for solution \tens{h} (Fig. \ref{Fig-h-overlap}) shows less but more
inclusive topics.\footnote{Note that  \tens{h}-topics in Fig. \ref{Fig-h-overlap} cover a smaller part of the their network
than \tens{hd}-topics in Fig.\ \ref{Fig-hd-overlap}. Especially, a lot of papers about galaxies are
missing in Fig. \ref{Fig-h-overlap} (cf.\ next section).}

 Similar to clustering \tens{h}, clustering \tens{hd} reconstructs the thematic split of contributions to \textit{Physical Review D}. The two largest clusters dominated by particle physics are \tens{hd10} and \tens{hd11} (cf.\ Table \ref{Tab:terms-hd} and Fig.\,\ref{Fig-hd-overlap}). Again, these clusters are dominated by papers published in \textit{Phys.\,Rev.\,D} in issues with uneven numbers, which are presented by in total 7760 papers in the 2003--2010 direct-citation network. 7358 of these are full papers in \tens{hd10} or \tens{hd11} (4229 of 4990 full papers in \tens{hd10}, 3129 of 3716 full papers in \tens{hd11}).

Clusters obtained by any hierarchical clustering form a hierarchy where each cluster equals the union of its subclusters. In the poly-hierarchy of link communities the union of a community's subcommunities does in general not fill it totally.

\subsection{\textit{Terra incognita}}
\label{sec:ti}
In both networks we have regions which have no substructures in our solutions. 
In clustering solution \tens{h}, our algorithm constructed overlapping valid  communities in a region comprising 72\,\% of the 2010 network, while the remaining 28\,\% were covered by just one community---the largest one (\tens{h1}) which covers 74\% of all links in the network. 

If we neglect the two largest communities \tens{hd1} and \tens{hd2}, clustering solution \tens{hd} covers 96\,\% of the network 2003--2010 but the third largest community, \tens{hd3}, has not any subtopic in the set of remaining 110 \tens{hd}-communities. If we unite all but the largest three \tens{hd}-communities the union covers only  68\,\% of all links.\footnote{There are 21 subtopics of \tens{hd3} among all 381 valid \tens{hd}-communities, but we have deselected them because their $\Psi$-value is above the threshold. Thus, \tens{hd3} has substructures but only faint ones and their inclusion would increase coverage only by 4\,\% to 72\,\%.} 
The results in both networks are consistent.\footnote{We found the area of clustering solution \tens{h} that is not occupied by communities but by \tens{h1} largely coinciding with the 2010 proportion of the area not covered by clustering solution \tens{hd} without the three largest communities (90\,\% of 3013 papers with membership $> 1/2$ not covered by the reduced  \tens{h}-solution are also in the set of 3881 papers not covered by the reduced \tens{hd}-solution).} 

Thus, we wonder why we have a topic with several thousand papers for which no subtopic can be found whereas for other smaller topics we find subtopics. 
The assumption that the unstructured region might represent thematically
diffuse areas that are produced by overlaps with neighbouring disciplines, could not be confirmed. 
For this to be the case, the area of the network would have to be of lower density, and publications would be expected to be located in those clusters of 
solution \tens{sr} that have a large share of non-Astronomy papers \cite[this issue]{Boyack2016Investigating}. Neither is the case.
In both networks the density of the \textit{terra incognita} is above the average density of the network.

Our algorithm could not identify thematic substructures in the \textit{terra incognita}. This observation is difficult to interpret. The other highly aggregated solutions presented in this special issue always place one cluster on the \textit{terra incognita} rather than assigning its papers to different clusters. However, solutions produced with high resolution levels and the first run of the Infomap algorithm construct smaller clusters in this region of the network (see also below, \ref{sec:Theresa2010}). This poses the question whether the cost function $\Psi$ evaluates certain regions of a network in a way that prevents the identification of substructures.

\subsection{Comparing the 2010 solution to a hard clustering of the same network}
\label{sec:Theresa2010}

\begin{table}[!t]
\caption{Link-based similarity indicators of the 14 \tens{uh}-clusters (size measured with number of papers), their best matched \tens{h}-communities (size measured as sum of paper fractions, cf.\ text), and keywords of \tens{uh}-clusters (cf.\ Table \ref{Tab:terms-h})}
\label{Tab:match.2010} 
\begin{tabular}{rrrrrrl}
\hline\noalign{\smallskip}
  & size&  & sizes  	& Salton& overlap &selected   \\ 
\tens{uh}& \tens{uh}& \tens{h}&  \tens{h}& index& min	  &keywords in \tens{uh}\\
\noalign{\smallskip}\hline\noalign{\smallskip}
1&  2418&	1& 10667.77& 	0.53& 1.00&galaxies, redshift, star formation\\
2&  1963&	2& 4798.45& 	0.53& 0.92&extrasolar planets, giant stars\\
3&  2508&	6& 2747.10& 	0.90& 0.97&background radiation, relativity\\
4&  1261&	8& 1835.36& 	0.70& 0.88&molecular clouds, protostars\\ 
5&  1548&	7& 2499.23& 	0.67& 0.86&neutron stars, gamma rays, pulsars\\
6&  1613&	10& 1321.53& 	0.87& 0.96&beyond the standard model\\
7&  1309&	12& 1360.39& 	0.95& 0.98&solar physics, stellar structure\\
8&  538&	1& 10667.77&	0.19& 0.97&white dwarfs, supernovae\\
9&  1103&	14& 1282.21&	0.87& 0.93&solar system, mars, natural satellites\\
10& 331&	7&  2499.23&	0.33& 0.97&gravitational waves,  astroparticle\\
11& 159&	50& 33.53&	0.66& 0.98&troposphere, mesosphere\\
12& 3&		1&  10667.77&	0.01& 0.96&galaxies, mass distribution\\
13& 9&		105& 8.50& 	0.99& 1.00&microgravity, earth orbit, the moon\\
14& 7&		107&	 7.00&	0.96& 1.00&celestial coordinate systems\\
\noalign{\smallskip}\hline
\end{tabular}
\end{table}

Theresa Velden applied the Infomap algorithm   to the network of 2010 papers and cited sources 
{(s.\ \shortciteN[this issue]{Velden2016Infomap} for a description of the approach). This provides us with the opportunity to compare our solution \tens{h} of 127 overlapping clusters to the solution \tens{uh}  of 14 hard clusters produced by the final iteration of the Infomap algorithm and also to the solution  produced by the first iteration of the Infomap algorithm with more than 1500 clusters.}

Table \ref{Tab:match.2010} provides the best matching \tens{h}-community
for each of the 14 \tens{uh}-clusters  (matches measured with link sets). 
The three \tens{uh}-clusters for which the match to communities from \tens{h} is worst (8, 10, and 12) are also those with highest values of cost function $\Psi$ (s.\ Fig.\ \ref{Fig-cost-size}, p.\ \pageref{Fig-cost-size}), which explains why our algorithm has not found good valid communities similar to them.

99\,\% of links of the large cluster \tens{uh1} are  in our \textit{terra incognita} (s.\ section \ref{sec:ti}) and \tens{uh1} covers 73\,\% of its links. 
{Thus, solution \tens{uh} has no substructures in this region of the network, either. However, the first iteration of the Infomap algorithm did find (small) thematic substructures in this region of the network. We are currently investigating the reasons for this discrepancy.}

\section{Discussion and Conclusion}
\label{sec:discussion}

The algorithm we developed and used constructs link communities by evolving populations of subgraphs in runs that are independent of each other. 
It does not enforce a compromise between possible different assignments of links or papers to clusters because it neither allocates all links or papers to clusters simultaneously, nor does it allocate links or papers exclusively.
Evolutions terminate depending on the cost function regardless of cluster sizes. For these reasons, we assume that the memetic algorithm is at least in principle able to reconstruct the ground truths we derived from our theoretical considerations.

Both our experiments produced poly-hierarchies of partially and completely overlapping valid communities, which we consider as representations of topics. 
Since the algorithm constructs each community independently from all others, it could also have constructed disjunct communities.  
Since it did not, overlapping topics appear to exist in networks of publications from the perspective of our evaluation function $\Psi$. 
This applies to both the network of 2010 papers and their cited sources and the direct citation network of 2003--2010 papers.

Since our algorithm could indeed construct overlapping communities, we can also conclude that it can serve the purpose of exploring theoretically derived structural properties of topics. 
This leads to the theoretically interesting question what these overlaps actually mean. The traditional hierarchical understanding of topics is that a larger topic consists of smaller topics. The overlapping topics we constructed suggest two questions that challenge this view. First, it seems possible that the smaller topics do not completely cover the larger one. Does this mean that there are publications that belong to a topic but to none of its subtopics? One can go even further by asking whether overlaps always represent subtopics in the sense of a more specific sub-area. Is it possible that a smaller topic overlapping a larger one is thematically different, e.g.\ by referring to a method applied only by few researchers working on the subject?

A downside of our approach is that it produces a large number of communities and thus forces two explicit decisions on the validity of communities, i.e.\ the representation of topics by communities. 
\textit{First}, we must decide how small a community can be and still represents a topic. This creates a sorites-type problem. One citation link between a paper and a cited source certainly does not constitute a topic. Neither do two citation links. Neither do three. What about four? Five? And so on. We must introduce a minimum size threshold for communities to be considered as representing topics, which admittedly is completely arbitrary. Comparisons to other solutions that produce very small clusters and experiments with varying size thresholds are necessary to provide some firmer ground for this decision.

\textit{Second}, we are forced to decide how different two communities must be in order to represent different topics. All approaches that identify communities by finding local minima in a rough cost landscape and allow for communities to overlap face the problem of selecting communities. A rough cost landscape has many local minima, which correspond to communities that share many nodes and links. This requires a numerical parameter (the resolution of the method) that defines the minimum number of links or nodes in which subgraphs must differ in order to be considered as different communities. We had to introduce such a parameter (cf. section \ref{sec:Psi}). Since the experiments reported here are the first ones with a new cost function, we set the parameter arbitrarily at 1/3 of the number of links of the community with higher $\Psi$-value. However, we expect further applications to work with a parameter whose value is derived from the purpose of the clustering exercise.

 Evolutionary algorithms feature many numerical parameters including population size, mutation variance, number of crossovers in each generation, and others. These technical parameters do not affect the content of the solution the same way as the resolution parameter. Instead, they affect the efficiency of the algorithm, i.e.\ the quality of a solution that is produced in a given computation time.
  
We presented a new algorithm and cost function which produce clustering solutions that meet our theoretical criteria and are comparable to those produced with other solutions (see \shortciteN{Velden2016Comparison} this issue on the comparison of solutions). However, the task of comprehensively validating the proposed approach lies still ahead of us. One of the common approaches to validation, the use of benchmark graphs, seems difficult to apply. To our knowledge, benchmark graphs whose topology reflects the ground truths our approach is supposed to reconstruct---thematic structures of scientific knowledge---do not yet exist. Since our approach is tailored to that task, testing its ability to simply detect communities in standard benchmark graphs seems beside the point.

The phenomenon of unstructured regions (section \ref{sec:ti}) needs further analysis.
In both clusterings there is a relatively large and dense region where 
neither we nor any other solution represented in this issue found
substructures.  
Is this a real phenomenon or is our cost function $\Psi$ unable to identify small topics in dense regions?
 
Although quite happy with the solutions the algorithm produces, we are much less happy with the computing costs of our memetic algorithm.\footnote{For example, one evolution in which a large community (with nearly 18\,\% of papers 2003--2010) is constructed run nearly six days on a machine with 16 CPUs.} To accelerate the procedure we already implemented parallel computing of populations in the R-Package PsiMin (cf.\ Appendix \ref{app:running}). We also run parallel evolutions on different machines and CPUs. Exploring the algorithm by systematically varying parameters requires experiments with smaller networks. Using one of the larger communities constructed in the experiments described in this paper could also contribute to further exploring comparative clustering solutions.

\bibliographystyle{chicago}

\begin{acknowledgements}
First, we want to thank all other authors of this special issue for doing this collaborative clustering exercise with us and for giving us many suggestions and hints. Special thanks to the members of the advisory board of our diversity project (\url{http://141.20.126.172/~div/}) who accompanied it throughout seven years. 
The work published here was partly funded by the German Research Ministry (01UZ0905). We thank Andreas Prescher for programming a fast C++-based R-package for parallel node-wise memetic search in the $\Psi$-landscape and a package for parallel link-wise local search.
We thank Theresa Velden for the giant component of the network of direct citations between papers published 2003--2010 and her cluster solution for the network of papers published 2010 obtained with Infomap. 
Thanks to Nees Jan van Eck for the cluster solutions for this network he had obtained with CitNetExplorer. We also would like to thank Kevin Boyack for additional data of his clustering. Kevin Boyack, Rob Koopman, and Shenghui Wang  have determined keywords we used to characterise clusters and communities. {We are grateful to Wolfgang Gl{\"a}nzel and Rob Koopman who commented on an earlier version of this paper, and 
to two anonymous reviewers whose critical remarks helped us to clarify our argument.}
\end{acknowledgements}

\bibliography{networks}   

\appendix

\section{Elements of Memetics}
\label{app.mem}

In large networks exploring the cost landscape by adding or removing individual links is very time-consuming. Therefore, we begin the search with a \textit{coarse} search phase in which the memetic algorithm adds or removes groups of links by adding or removing a node with all its links. Node-wise memetics is done with subgraphs induced by node sets $C$: mutation, crossover, and local search are based on node sets. The coarse search phase is followed by a \textit{fine} search phase, namely a link-wise local search (cf.\ Algo\-rithm~\ref{pseudcode-exp}).
\label{node-wise}

\paragraph{Local search:}
Local search (adaptation) in the cost landscape is done by a greedy algorithm for finding local cost minima that correspond to communities. 
The algorithm is called greedy because it always chooses the step in the cost landscape that brings the biggest decrease (or the smallest increase)  of $\Psi$. 
Node-wise local search includes a neighbouring node of subgraph $S$ (or excludes a boundary node)
with all its links to nodes in $S$.  
Link-wise local search includes a neighbouring link of subgraph $S$ or excludes a link attached to a boundary node.
Local search can begin by a series of either inclusions or exclusions of nodes (links). 
If no neighbouring place in the landscape with lower cost can be reached the algorithm excludes or includes further nodes (links). This stops after a maximum number of steps without improvement. This maximum number is calculated from the relative resolution parameter. In all experiments described here we go as many steps as allowed by the rule that a community's range should be at least one third of its size \shortcite[cf.\ p.\ 6]{havemann_detecting_2015}.
When no further improvement can be achieved, the search switches from inclusion to exclusion or vice versa. 
Inclusion and exclusion are continued until no further improvement is possible. 
Exclusion can make the subgraph unconnected. Node-wise local search then proceeds with its main component. In the case of link-wise local search the exclusion process can go through
places in the cost landscape which correspond to unconnected subgraphs. Connectedness is tested only after running the algorithm. If we obtain two or more components then we start new link-wise local searches using components as seeds.

\paragraph{Mutation:}
We mutate a community $C$ with mutation variance $v < 1$ by changing  maximally a 
proportion $v$ of its nodes. This is done by randomly selecting a node in $C$, to which neighbours in $C$ are randomly added until a connected subgraph of a size larger than $(1 - v)|C|$ is reached. To this subgraph, further random neighbours are added, which now can also be outside $C$. This random extension continues until a connected subgraph of size $|C|$ is reached. The mutant is then subjected to two local searches, one each starting with greedy inclusion and exclusion of nodes. Thereby 
we obtain two communities from each mutation (which can be identical). 

\paragraph{Crossover:}
From two parent communities we construct two new individuals by taking intersection and union of the communities as starting points for adaptive local searches. Adapting the union is started with exclusion of nodes, adapting the intersection is started with inclusion of nodes. Of course, it has no effect to cross such parents where one of them is part of the other one. We also do not cross disjoint parents because we are interested in connected subgraphs as solutions. 

\paragraph{Selection:}
From the old population and the results of mutations and crossovers we select the communities with lowest $\Psi$-values, keeping the population size constant. A new best  community is only included if it is inside the minimal range (defined by the resolution parameter) of the  best community of the original population.
Deselected new best communities can be used as seeds for other memetic searches. 

\section{Running memetic search}
\label{app:running}

\label{running4steps} \label{running}
Each seed was first adapted by a local search and then used to initialise the population of 16 different communities by mutating the adapted seed with a variance of 15\,\%.  

In the first round of memetic searches, up to five experiments were run with each seed. The standard mutation variance in each experiment was 2\,\%, i.e.\ up to 2\,\% of the nodes were randomly exchanged. The variance was increased to 15\,\%  if $\Psi$-values did not improve for 10 generations (renewal of population). For each seed, the results of these experiments where used to initialise a population (of eight communities) for a second round of up to ten memetic experiments. Here we omitted renewal and led the mutation variance decrease after each generation for which no improved best community was obtained. This strategy was applied to find nearby better communities which could have been overseen with fixed mutation variance. After node-wise memetics, for each community we made a link-wise local search. 

\label{sec:post}
We tested whether link communities have a range above the minimum range of one third of the link set's size. At this stage, 
we also considered subgraphs induced by complementary link-sets because cost function $\Psi(L)$ is the same for $L$ and its complement. If the complement is unconnected 
then we tried to improve its main component by link-wise local search.

We implemented the whole procedure as R-scripts. For parallel node-wise memetic search in the $\Psi$-landscape we applied a fast new C++-based  R-Package PsiMin. For parallel link-wise local search we applied R-Package PsiMinL. Both (yet unpublished) packages
were programmed for us by Andreas Prescher.  

\section{Results}
\label{appendix}

\begin{table}[!p]
\caption{Size data, cost, numbers of subtopics, of supertopics, and of other related topics of 50 largest link clusters 2010 (clustering \tens{h}, relation data without \tens{h1})}
\begin{tabular}{rrrr rrrr }
\hline\noalign{\smallskip}
 \tens{h} =   &  $|L| =$ &   & \# full & sum of &  \# related  &&  \\
$|L|$-rank & \# links &  $\Psi(L)$ & papers & fractions & subtopics &  supertopics &  others  \\
\noalign{\smallskip}\hline\noalign{\smallskip}
\label{tab:2010} 
1 & 394924 & 0.0391 & 10354 & 10667.77 & -- & -- & --\\
2 & 183177 & 0.0605 & 5741 & 5934.75 & 70 & 0 & 34\\
3 & 141063 & 0.0391 & 3923 & 4095.23 & 24 & 0 & 32\\
4 & 111323 & 0.0708 & 3518 & 3650.07 & 47 & 1 & 46\\
5 & 93641 & 0.0713 & 3203 & 3299.33 & 61 & 1 & 39\\
6 & 91349 & 0.0550 & 2604 & 2747.10 & 17 & 1 & 31\\
7 & 74369 & 0.0967 & 2354 & 2499.23 & 28 & 0 & 61\\
8 & 63508 & 0.0967 & 1738 & 1835.36 & 0 & 1 & 48\\
9 & 47464 & 0.0348 & 1907 & 1964.06 & 32 & 1 & 44\\
10 & 44729 & 0.0415 & 1272 & 1321.53 & 2 & 1 & 30\\
11 & 35259 & 0.0821 & 1180 & 1276.35 & 7 & 0 & 32\\
12 & 31999 & 0.0336 & 1321 & 1360.39 & 20 & 2 & 46\\
13 & 31232 & 0.0475 & 974 & 1008.05 & 7 & 1 & 16\\
14 & 27685 & 0.0405 & 1236 & 1282.21 & 40 & 3 & 47\\
15 & 24304 & 0.0912 & 823 & 890.08 & 4 & 3 & 21\\
16 & 22111 & 0.0638 & 1109 & 1176.34 & 52 & 3 & 43\\
17 & 19343 & 0.0890 & 724 & 762.28 & 5 & 2 & 18\\
18 & 19161 & 0.0565 & 589 & 621.82 & 1 & 3 & 15\\
19 & 19058 & 0.0417 & 841 & 874.71 & 28 & 5 & 42\\
20 & 18036 & 0.0640 & 723 & 762.81 & 19 & 3 & 31\\
21 & 14711 & 0.0460 & 702 & 737.56 & 25 & 6 & 53\\
22 & 13244 & 0.0636 & 728 & 754.02 & 43 & 2 & 44\\
23 & 13177 & 0.0899 & 494 & 512.70 & 6 & 2 & 13\\
24 & 13079 & 0.0776 & 679 & 698.55 & 29 & 2 & 42\\
25 & 13008 & 0.0826 & 665 & 688.18 & 19 & 3 & 46\\
26 & 12540 & 0.0630 & 403 & 414.81 & 0 & 4 & 9\\
27 & 8233 & 0.0609 & 455 & 470.90 & 31 & 1 & 43\\
28 & 6982 & 0.0861 & 223 & 235.07 & 3 & 2 & 18\\
29 & 6023 & 0.0424 & 239 & 245.11 & 6 & 8 & 31\\
30 & 5812 & 0.0814 & 387 & 399.25 & 28 & 4 & 44\\
31 & 5771 & 0.0889 & 230 & 242.44 & 0 & 6 & 30\\
32 & 5734 & 0.0923 & 185 & 196.42 & 1 & 1 & 17\\
33 & 5593 & 0.0688 & 328 & 343.54 & 16 & 7 & 39\\
34 & 5446 & 0.0853 & 195 & 210.40 & 2 & 4 & 8\\
35 & 4441 & 0.0664 & 307 & 314.62 & 16 & 8 & 41\\
36 & 3999 & 0.0879 & 253 & 259.45 & 6 & 4 & 43\\
37 & 3755 & 0.0945 & 194 & 202.32 & 12 & 7 & 23\\
38 & 3580 & 0.0770 & 123 & 135.36 & 2 & 5 & 6\\
39 & 3484 & 0.0944 & 142 & 154.64 & 1 & 8 & 31\\
40 & 2750 & 0.0690 & 165 & 169.76 & 5 & 12 & 39\\
41 & 2204 & 0.0811 & 99 & 101.35 & 6 & 5 & 38\\
42 & 1816 & 0.0331 & 132 & 137.73 & 4 & 5 & 30\\
43 & 1316 & 0.0519 & 134 & 139.33 & 8 & 15 & 28\\
44 & 1290 & 0.0502 & 105 & 108.67 & 7 & 11 & 22\\
45 & 1183 & 0.0888 & 58 & 59.00 & 3 & 8 & 30\\
46 & 1072 & 0.0702 & 64 & 64.82 & 7 & 5 & 33\\
47 & 925 & 0.0470 & 83 & 85.91 & 5 & 7 & 28\\
48 & 889 & 0.0558 & 74 & 76.04 & 1 & 20 & 13\\
49 & 789 & 0.0710 & 30 & 30.41 & 1 & 9 & 15\\
50 & 709 & 0.0325 & 32 & 33.53 & 1 & 6 & 12\\

\noalign{\smallskip}\hline
\end{tabular}
\end{table}

\begin{table}[!p]
\caption{Size data, cost, numbers of subtopics, of supertopics, and of related topics of 50 largest link clusters 2003--2010 (clustering \tens{hd}, relation data without \tens{hd1} and \tens{hd2})}
\label{tab:8y} 
\begin{tabular}{rrrr rrrr }
\hline\noalign{\smallskip}
 \tens{hd} =   &  $|L| =$ &   & \# full & sum of &  \# related  &&  \\
$|L|$-rank & \# links &  $\Psi(L)$ & papers & fractions & subtopics &  supertopics &  others  \\
\noalign{\smallskip}\hline\noalign{\smallskip}
1 & 661025 & 0.0621 & 62606 & 67817.87 & -- & -- & --\\
2 & 611025 & 0.0824 & 71488 & 79068.49 & -- & -- & --\\
3 & 313717 & 0.0824 & 19948 & 22756.18 & 0 & 0 & 24\\
4 & 263676 & 0.0621 & 31242 & 33989.63 & 53 & 0 & 16\\
5 & 138567 & 0.0549 & 17372 & 18875.51 & 41 & 0 & 59\\
6 & 131062 & 0.0920 & 11899 & 13664.33 & 0 & 0 & 42\\
7 & 66007 & 0.0634 & 10406 & 11219.33 & 52 & 1 & 14\\
8 & 51956 & 0.0240 & 9035 & 9599.80 & 19 & 0 & 38\\
9 & 34292 & 0.0391 & 8201 & 8811.90 & 48 & 2 & 18\\
10 & 27006 & 0.0230 & 4990 & 5299.93 & 13 & 0 & 14\\
11 & 21283 & 0.0672 & 3716 & 4161.44 & 4 & 0 & 30\\
12 & 20617 & 0.0846 & 2740 & 3049.89 & 5 & 0 & 35\\
13 & 15851 & 0.0377 & 4846 & 5183.65 & 32 & 3 & 26\\
14 & 15202 & 0.0603 & 2789 & 3146.62 & 6 & 1 & 29\\
15 & 11787 & 0.0472 & 2919 & 3175.17 & 10 & 4 & 29\\
16 & 10840 & 0.0331 & 3873 & 4109.32 & 33 & 4 & 22\\
17 & 7726 & 0.0699 & 1203 & 1381.78 & 3 & 1 & 17\\
18 & 7027 & 0.0675 & 1254 & 1412.50 & 0 & 1 & 12\\
19 & 5919 & 0.0582 & 923 & 1057.08 & 0 & 1 & 20\\
20 & 5898 & 0.0788 & 1312 & 1482.75 & 4 & 1 & 24\\
21 & 5519 & 0.0755 & 1023 & 1159.48 & 0 & 1 & 16\\
22 & 5306 & 0.0524 & 1715 & 1928.48 & 12 & 3 & 25\\
23 & 5130 & 0.0700 & 651 & 746.04 & 0 & 2 & 7\\
24 & 5039 & 0.0350 & 1907 & 2036.18 & 13 & 5 & 29\\
25 & 4607 & 0.0156 & 2129 & 2214.93 & 22 & 2 & 26\\
26 & 4585 & 0.0680 & 743 & 858.58 & 0 & 1 & 17\\
27 & 3734 & 0.0776 & 876 & 1001.57 & 2 & 2 & 23\\
28 & 3558 & 0.0426 & 1120 & 1210.82 & 5 & 8 & 14\\
29 & 3052 & 0.0920 & 670 & 775.94 & 2 & 0 & 28\\
30 & 2826 & 0.0729 & 834 & 944.41 & 3 & 1 & 26\\
31 & 2613 & 0.0879 & 413 & 481.93 & 0 & 5 & 14\\
32 & 2461 & 0.0239 & 1042 & 1108.51 & 8 & 8 & 18\\
33 & 2413 & 0.0627 & 769 & 847.31 & 1 & 9 & 9\\
34 & 1990 & 0.0292 & 996 & 1058.81 & 7 & 3 & 36\\
35 & 1883 & 0.0572 & 344 & 395.04 & 0 & 1 & 13\\
36 & 1806 & 0.0165 & 941 & 978.65 & 7 & 3 & 29\\
37 & 1473 & 0.0834 & 510 & 585.49 & 5 & 1 & 18\\
38 & 1415 & 0.0373 & 657 & 711.38 & 5 & 3 & 26\\
39 & 1259 & 0.0731 & 251 & 291.21 & 0 & 6 & 12\\
40 & 1230 & 0.0853 & 308 & 351.10 & 1 & 4 & 16\\
41 & 1171 & 0.0905 & 458 & 507.82 & 3 & 6 & 18\\
42 & 1143 & 0.0889 & 321 & 366.07 & 0 & 1 & 23\\
43 & 1120 & 0.0671 & 170 & 198.11 & 0 & 1 & 16\\
44 & 1117 & 0.0761 & 360 & 398.15 & 1 & 3 & 21\\
45 & 1066 & 0.0830 & 318 & 366.26 & 2 & 1 & 10\\
46 & 1045 & 0.0230 & 521 & 548.00 & 1 & 11 & 17\\
47 & 971 & 0.0729 & 236 & 272.84 & 0 & 3 & 12\\
48 & 937 & 0.0492 & 508 & 548.67 & 5 & 4 & 28\\
49 & 906 & 0.0612 & 228 & 263.50 & 1 & 1 & 15\\
50 & 803 & 0.0516 & 417 & 455.16 & 2 & 4 & 30\\
\noalign{\smallskip}\hline
\end{tabular}
\end{table}

In Table \ref{tab:2010} and Table \ref{tab:8y} we list data of the 50 largest communities in the networks with papers published in 2010 and in 2003--2010, respectively. The following definitions are used: 

\paragraph{Full papers} are papers which have no citation links to papers outside the subgraph induced by link set~$L$. 

\paragraph{Fractions of papers} are membership grades of papers i.e.\ the ratios of numbers of internal to numbers of all citation links. Thus, \textit{sum of fractions} is determined as $$\sum_{i=1}^n \frac{k_i^\mathrm{in}(L)}{k_i},$$
where $k_i$ is the degree of paper $i$ and $k_i^\mathrm{in}(L)$ its internal degree with regard to $L$ (the number of links attached to $i$ which are in link set~$L$).

\paragraph{Subtopics} of community $L$ are all other communities which have at least 95\,\% of their links in link set~$L$. 

\paragraph{Supertopics} of community $L$ are all other communities which include at least 95\,\% of links in link set~$L$.

\paragraph{Other related  communities}  of community $L$ are all those communities  which have links in common with link set $L$ but are not sub- or supertopics of~$L$. 

\end{document}